\documentclass[12pt,emtex]{article}
\usepackage{amsfonts}
\usepackage{amssymb}
\usepackage{amscd}
\usepackage{amsmath}
\newcommand{\BEQ}{\begin{equation}}
\newcommand{\EEQ}{\end{equation}}

\newtheorem{Th}{Theorem}

\newtheorem{Lem}{Lemma}
\newtheorem{Prop}{Proposition}
\newtheorem{Cor}{Corollary}

\newtheorem{Ex}{Example}
\newtheorem{Def}{Definition}
\newtheorem{Rem}{Remark}

\def\nn{\nonumber}

\def\bea{\begin{eqnarray}}
\def\eea{\end{eqnarray}}

\def\S{{\Sigma}}
\def\O{{\cal O}}
\def\C{{\mathbb{ C}}}
\def\CC{{\mathbb{ C}}}
\def\Cz{{\mathbb{ C}}[z]}
\def\ep{{\epsilon}}
\def\ML{M_\Lambda}

\def\F{{\cal F}}
\def\K{\mathcal{K}}
\def\L{\Lambda}

\begin{document}

\begin{titlepage}
\hfill ITEP-TH-82/02 \vskip 2.5cm

\centerline{\LARGE \bf Hitchin systems on singular curves II. }
\centerline{\LARGE \bf Gluing subschemes. }

\vskip 1.0cm \centerline{A. Chervov \footnote{E-mail:
chervov@itep.ru} , D. Talalaev \footnote{E-mail:
talalaev@gate.itep.ru} \footnote{The work of the both authors
has been partially supported by the RFBR grant 01-01-00546,
the work of A.C. was partially supported by the
Russian President's grant 00-15-99296, the work of D.T.
has been partially supported by the RFBR grant for the support
of young scientists MAC-2003 N 03-01-06236 under the
project 01-01-00546.
}}

\centerline{\sf Institute for Theoretical and Experimental Physics
\footnote{ITEP, 25 B. Cheremushkinskaya, Moscow, 117259, Russia.}}

\vskip 2.0cm

\centerline{\large \bf Abstract}

\vskip 1.0cm

In this paper we continue our studies of Hitchin systems on singular curves
(started in hep-th/0303069).
We consider a rather general class of curves which can be obtained from the projective line
by gluing  two subschemes together (i.e. their  affine part is:
Spec $\{f \in \CC[z]: f(A(\ep))=f((B(\ep)); \ep^N=0 \},$
where $A(\ep), B(\ep)$  are arbitrary polynomials) .
The most simple examples are the generalized cusp curves which are projectivizations of
Spec $\{f \in \CC[z]: f'(0)=f''(0)=...f^{N-1}(0)=0 \}$).
We describe the geometry of such curves;
in particular we calculate their genus
(for some curves the calculation appears to be related with the
iteration of polynomials $A(\ep), B(\ep)$
defining the subschemes). We obtain the explicit description
of moduli space of vector bundles, the  dualizing sheaf, Higgs field and
other ingredients of  the Hitchin integrable systems;
these results may deserve the independent interest.
We prove the integrability of Hitchin systems on such curves.
To do this we develop $r$-matrix formalism
for the functions on the
truncated loop group $GL_n(\CC[z]), z^N=0$.
We also show how to obtain the Hitchin integrable systems on such curves as
hamiltonian reduction from the more simple system on some finite-dimensional
space.


\end{titlepage}

\tableofcontents

\section{Introduction}

In this paper we continue our studies of Hitchin systems on singular curves.
Rational singular curves provide large class of explicit, but nontrivial
examples of the   description of the Hitchin system and all their
ingredients: the
moduli space of vector bundles, the dualizing sheaf, Higgs field etc.
Such explicit examples are quite important, because
Hitchin system, despite its importance, is far from being
fully investigated. In such examples one can hope to
work out methods for complete understanding of the subject.

In the previous paper \cite{CT1} we have considered
the Hitchin system on
rational singular curves which can be obtained from the projective line
by gluing several points together or by taking
cusp singularities.
The main idea of the present paper that
the more general case of curves obtained by  gluing subschemes
can be treated along the same lines as it was done in the previous paper.
So we largely extend the class of examples where
Hitchin system can be explicitly described.

We hope that one can read this paper independently from
\cite{CT1}, though sometimes it may be useful for the reader
 to look in it. We also assume  the reader to be  familiar
with Hitchin's original paper \cite{Hit}.

However we recall here some principal steps of the construction.
Original Hitchin system lives on the cotangent bundle to the space
$\mathcal{T}^*\mathcal{M}$ of
stable holomorphic bundles on nonsingular algebraic curves $\S.$
A point of the phase space corresponds to the pair $(E,\Phi)$ where $E$ is a
holomorphic bundle and $\Phi$ is a cotangent vector to the moduli space. By
standard arguments from deformation theory the tangent vector to the moduli
space at the point $E$ can be identified with an element of
$H^1(\S,End(E)).$ The Serre's pairing provide a geometric description of
cotangent vectors, indeed, a cotangent vector $\Phi$ can be identified with an
element of $H^0(\S, End(E)\otimes\mathcal{K})$ where $\mathcal{K}$ is the
canonical class of $\S.$ One proceeds by constructing the following dynamical
system which turns out to be integrable: one takes the canonical symplectic
structure on the cotangent bundle and the family of functions $I_{k,l}$
which are the coefficients obtained by expanding  $Tr\Phi^k(z)$ on the
basic of $H^0(\mathcal{K}^k).$

In this paper we continue to generalize this
construction to the case of singular algebraic curves.
In order to emphasize the analogy between the
case of gluing points and   subschemes,
let us at first  recall some of the  results from \cite{CT1},
after that we describe the results of the present
paper.

\begin{itemize}
\item Consider the curve
$\Sigma^{proj}$ which results from gluing $2$
distinct points $P_i,i=1,2$ on $\CC
P^1$ to one point (i.e. the curve which is obtained by adding the smooth
point $\infty$ to the curve $ \Sigma^{aff}= Spec \{ f\in \CC[z]:
~ f(P_1)=f(P_2)  \}$).
  \item A rank $r$ vector bundle on such a curve
corresponds to a rank
$r$  module $\ML$ over the affine part given by the subset of vector-valued
functions $s(z)$ on $\CC$
i.e. $s(z)\in \CC[z]^r$ which satisfy the conditions: $s(P_1)=\L s(P_{2})$.
The moduli space of vector bundles on $\Sigma^{proj}$ is the
factor by $GL_r$ of the set of invertible matrices $\L$ where
$GL_r$ acts by conjugation.
  \item The space of global
sections of the dualizing sheaf on $\Sigma^{proj}$  is one-dimensional
and basic section can be described as
meromorphic
differential on $\CC$ given by  $\frac{dz}{z-P_1} - \frac{dz}{z-P_2} $.
  \item The
endomorphisms of the module $\ML$ are matrix valued polynomials $\Phi(z)$
such that $\Phi(P_1)=\L \Phi(P_2)\L^{-1}$. The action of $\Phi(z)$ on $s(z)$
is: $s(z)\mapsto \Phi(z)s(z)$. The space $H^1(End(\ML))$ can be described as
the space $gl[z]$ of matrix valued polynomials factorized by the subspaces:
$End_{out}=\{\chi(z)\in gl[z]|\chi(z)=const\}$ and $End_{in}=\{\chi(z)\in
gl[z]|\chi(P_1)=\L \chi(P_2)\L^{-1} \}$. The elements of $H^1(End(\ML))$ are
the tangent vectors to the moduli space of vector bundles at the point
$\ML$. The  element $\chi(z)$ gives the following deformation of $\L$:
        \bea
        \delta_{\chi(z)} \L = \chi(P_1)\L-\L\chi(P_2)
        \eea
  \item The global sections of $H^0(End(\ML)\otimes \K)$ ("Higgs fields")
are described as
\bea
\Phi (z)=
\frac{\L \Phi\L^{-1} }{z-P_1}dz
 -\frac{\Phi}{z-P_2}dz, \label{Lax}
\eea
where $\L \Phi\L^{-1}-\Phi=0$.
\item The symplectic form on the cotangent bundle to the moduli space
can be described as the reduction of the form on the space $\L,\Phi$
given by
\bea
 Tr d(\L^{-1} \Phi )   \wedge d \L \label{sf1}.
\eea

  \item  The hamiltonians
 $Tr (\Phi(z)^k)$ Poisson
 commute  on the non-reduced phase space
 with each  other for any $k,z$,
 hence, due to their invariance,
 gives commuting family of hamiltonians
 on the reduced phase space.

\end{itemize}

{\bf Remarks:}
For these particular  case of gluing two points the same Lax operator
$\Phi (z)$
has been
proposed by N. Nekrasov (\cite{NN}), though his methods are different from
ours, and the explicit description of bundles, dualizing sheaf,
endomorphisms etc are absent in his approach.

In the present paper instead of a points $P_i\in \CC$
we consider a  subschemes $A,B\in \CC$.
Algebraically the point $P$ is described as homomorphism
$\CC[z]\to \CC$ given by $f\mapsto f(P)$.
Analogously one can describe subschemes as
homomorphisms of rings with units
 $\CC[z]\to \CC[\ep]$, where $\ep^N=0$.
Such homomorphisms are uniquely defined by the
image of $z$ in $\CC[\ep]$, which can be arbitrary polynomial
$A(\ep)\in \CC[\ep]$.

{\bf Notation.} In this paper we will denote by subscheme $A(\ep)$,
where $A(\ep)$ is an arbitrary polynomial, the subscheme
in $\CC$ or $\CC P^1$ which are defined by the
homomorphisms $\phi: \CC[z]\to \CC[\ep]$, given by
$\phi: z \mapsto A(\ep)\in \CC[\ep]$, where $\ep^N=0$.

{\bf Notation.} The number $N$ always denote exponent
such that $\ep^N=0$.

Let us summarize the main results of the present paper
(in the main text we consider more general examples,
but here we cite most illustrative ones).

\begin{itemize}
\item Consider the curve
$\Sigma^{proj}$ which results from gluing $2$ arbitrary
subschemes
$A(\ep), B(\ep)$  on $\CC
P^1$ to one point (i.e. the curve which is obtained by adding the smooth
point $\infty$ to the curve $\Sigma^{aff}= Spec \{ f\in \CC[z]:
~ f(A(\ep))=f(B(\ep)) \}$, where $\ep^{N}=0$ ).
In our paper  for some $A(\ep), B(\ep)$
 we calculate the genus (i.e. $dim H^1(\O)$)
of such curves (see section \ref{sect-desrpt-curves}
proposition \ref{prop-curves}), which nontrivially depends
on $A(\ep), B(\ep)$.
The basic examples to keep in mind are the following.
\begin{itemize}
\item Nilpotent: $A(\ep)=\ep, B(\ep)=0$, the genus equals $N-1$.
\item Root of unity: $A(\ep)=\ep, B(\ep)=\alpha \ep$,
where $\alpha^k=1$,
\\
 the genus equals $N-1-[(N-1)/k]$.

More generically one can consider:
$A(\ep)=\ep$ and $B(\ep)$ such that $\underbrace{B(B(B...(B}_{\mbox{k times}}
  (\ep)))=\ep$ $mod$
$\ep^{N-1}$ then the genus will be the same.

\item Different geometric points: $A(\ep)=a_0+a_1\ep+...
+a_{N-1}\ep^{N-1},
B(\ep)=b_0+b_1\ep+...+b_{N-1}\ep^{N-1}$, such that
$a_0\ne b_0$, the genus equals $N$.
\end{itemize}

\item Rank $r$ vector bundles on such a curve
correspond  to some of  rank
$r$  modules $\ML$ over the affine part given by the subset of vector-valued
functions $s(z)$ on $\CC$
i.e. $s(z)\in \CC[z]^r$
which satisfy the conditions: $s(A(\ep))=\L(\ep) s(B(\ep))$,
where $\L(\ep)=\sum_{i=0,...,N-1} \L_i \ep^i$ is matrix valued
polynomial. The conditions of  projectivity of the module
$\ML$ (and hence the  condition for corresponding
sheaf over $\CC P^1$ to be vector bundle) are
the following (see section \ref{proj-mod}
proposition \ref{prop-proj-mod}).
\begin{itemize}
\item Nilpotent: $A(\ep)=\ep, B(\ep)=0$, the condition is: $\L_0=Id$.
\item Root of unity: $A(\ep)=\ep, B(\ep)=\alpha \ep$,
where $\alpha^k=1$,
\\
the condition is:  $\L(\ep)\L(\alpha \ep)...\L(\alpha^{k-1} \ep) =Id$.
\item Different geometric points: $A(\ep)=a_0+a_1\ep+...
+a_{N-1}\ep^{N-1},
B(\ep)=b_0+b_1\ep+...+b_{N-1}\ep^{N-1}$, in
this case the only condition: $\L_0$ must be invertible.
\end{itemize}

 The moduli space of vector bundles on $\Sigma^{proj}$ is the
factor by $GL_r$ of the set of $\L(\ep)$ which satisfies
the conditions above, where $GL_r$ acts by conjugation
(see section  \ref{sect-vect-bund} theorem \ref{bund-on-sigm}).

  \item The  global
sections of the dualizing sheaf on $\Sigma^{proj}$ can be described as
meromorphic
differentials on $\CC$ given by (see
section  \ref{sect-dual-expl}
 proposition \ref{hol-dif-main}):
\bea
Res_{\ep} \bigl(\frac{\phi(\ep) dz}{z-A(\ep)} -
\frac{\phi(\ep) dz}{z-B(\ep)}\bigr),
\eea
where $\phi(\ep)=\sum_{i=0,...,N-1} \phi_i \frac{1}{\ep^{i+1}}$
is arbitrary. This expression should be understood
expanding the denominators in  geometric progression series:
\begin{equation*}
\begin{split}
\frac{1}{z-A(\ep)}&=\frac{1}{z-a_0-a_1\ep-a_2\ep^2-...}=
\frac{1}{(z-a_0)(1- \frac{a_1\ep+a_2\ep^2+...}{z-a_0})}\\
&= \frac{1}{(z-a_0)}{(1+ \frac{a_1\ep+a_2\ep^2+...}{z-a_0}
+(\frac{a_1\ep+a_2\ep^2+...}{z-a_0})^2 +... }),
\end{split}
\end{equation*}
and $Res_{\ep}$ means taking the coefficient at $\frac{1}{\ep}$.
Our claim is that for all $\phi(\ep)$ the expression
above gives global holomorphic differential on singular
curve $\Sigma^{proj}$ and all the differentials
can be obtained in such a way. In general
the map from $\phi(\ep)$ to holomorphic differentials
has a kernel.

  \item The
endomorphisms of the module $\ML$ are matrix valued polynomials $\Phi(z)$
such that
$\Phi(A(\ep))=\L(\ep) \Phi(B(\ep))\L(\ep)^{-1}$.
The action of $\Phi(z)$ on $s(z)$
is: $s(z)\mapsto \Phi(z)s(z)$. The space $H^1(End(\ML))$ can be described as
the space $gl[z]$ of matrix valued polynomials factorized by the subspaces:
$End_{out}=\{\chi(z)\in gl[z]|\chi(z)=const\}$ and $End_{in}=\{\chi(z)\in
gl[z]|\chi(A(\ep)))=\L(\ep) \chi(B(\ep))\L(\ep)^{-1} \}$.
The elements of $H^1(End(\ML))$ are
the tangent vectors to the moduli space of vector bundles at the point
$\ML$. The  element $\chi(z)$ gives the following deformation of $\L$:
        \bea
        \delta_{\chi(z)} \L(\ep) = \chi(A(\ep))\L(\ep)-\L(\ep)
        \chi(B(\ep)).
        \eea

  \item The global sections of $H^0(End(\ML)\otimes \K)$ ("Higgs fields")
are described as (see section
\ref{end-tens-k} proposition \ref{hol-dif-k}):
\bea
\Phi (z)=
Res_{\ep} \left( \frac{ \Phi(\ep) }{z-A(\ep)}dz
 -\frac{ \L(\ep)^{-1} \Phi(\ep) \L(\ep)}{z-B(\ep)}dz \right),
\label{Lax-ep}
\eea
where $Res_\ep \L(\ep) \Phi(\ep) \L(\ep) ^{-1}-\Phi(\ep)=0$;
and
$\Phi(\ep)= \sum_i \Phi_i \frac{1}{\ep^{i+1}}$ is a matrix
valued polynomial.
This expression should be understood
expanding the denominators in  geometric progression series,
as it was explained above. And we also claim that
all global sections from $H^0(End(\ML)\otimes \K)$ can be
obtained in such a way and for all $\Phi(\ep)$ the
expression above gives global
section from $H^0(End(\ML)\otimes \K)$.

\item
The symplectic form on the cotangent bundle to the moduli space
can be described as the restriction and reduction  of the form on the space
$\L(\ep),\Phi(\ep)$
given by (see section \ref{sect-1-form} theorem \ref{th-1-form}):
\bea
 Res_{\ep} Tr d( \L(\ep)^{-1} \Phi(\ep) )   \wedge d \L(\ep)
\label{sf1-ep}.
\eea


\end{itemize}

So saying shortly:

{\bf Result:} (see theorem \ref{red-th})
The Hitchin system on the curve $\Sigma^{proj}$
can be described as the system with a phase space which is the hamiltonian
reduction of the space of
$\L(\ep),\Phi(\ep)$
(more precisely we should speak about its subspace
defined by the conditions
mentioned in the second item).
Symplectic form is given by the formula \ref{sf1-ep}.  The
reduction is taken by the group $GL(r)$, which acts by conjugation.
The Lax operator is given by formula \ref{Lax-ep}. (Hence
hamiltonians are coefficients of the expansion of $Tr (\Phi(z)^k)$
at the basic of holomorphic $k$-differentials $H^0(\K^k)$).
Let us emphasize that $ \forall z,w,k,l $ it is true that
 $Tr (\Phi(z)^k)$ and
$Tr (\Phi(w)^l)$ Poisson commute on the {\it nonreduced} phase space.

In order to prove integrability of Hitchin system in case of our singular
curves one should prove that hamiltonians Poisson commute.
This is done in section \ref{commut-sect}  by use of $r$-matrix technique,
the propositions may be interesting by themselves so let us formulate them
also.

\begin{itemize}
\item The bracket between $\Phi(\ep)$ is of $r$-matrix
type (see lemma \ref{l1}):
\begin{equation} 
  \{\Phi(\ep)\otimes\Phi(\eta)\}=\left| [\Phi(\ep)\otimes
1,\delta_\eta^{\ep}R]\right|_{-}=\left|[\Phi(\ep)\otimes
1+1\otimes \Phi(\eta),\delta_{\eta}^{\ep}R]\right|_{\eta}.
\end{equation}
Where $\{A\otimes B\}$ is standard St.-Petersburg's
 notation for brackets between matrices meaning
that the result is a matrix in the tensor product of spaces with matrix elements
which are Poisson brackets between the matrix elements of matrixes $A,B$.
The matrix $R$ is just permutation matrix: $R(u\otimes v)=v\otimes u$.
The function
$\delta_\ep^\eta $ is equal to $\sum_{k=0}^{N-1} \frac {\eta^k}
{\ep^{k+1}}$  and $|...|_{\nu}$ is the truncation of a polynomial
i.e. taking it's part consisting of the monomials
$\nu^k$, where $k<N$.
The formula is easy to prove and it
 will be standard if there will be no truncations
and all sums will be infinite i.e. $N=\infty$.
\item
What is more surprising for us is that the following
brackets are also of $r$-matrix form (see lemma \ref{lem2}):
\bea
\left\{\Phi(z,\ep)\otimes\Phi(w,\eta)\right\}=\left|\left[\Phi(z,\ep)\otimes
1,R\right]\delta_{\eta}^{\ep}\delta_w^z\right|_{-\{\mbox{\tiny all
variables}\}}. \nn
\eea
Where $\Phi(z,\ep)=\left|\frac {\Phi(\ep)}{z-A(\ep)}-\frac
{\L^{-1}(\ep) \Phi(\ep) \L(\ep) }{z-B(\ep)}\right|_{-\{\ep\}}$.
\item
And finally the desired bracket between the Lax operators for Hitchin system
is also of $r$-matrix type (see lemma \ref{r-for-phiz}):
\begin{equation}\label{PB7-intro}
  \left\{\Phi(z)\otimes\Phi(w)\right\}=\left|[\Phi(z)\otimes
  1,R]\delta_w^z\right|_{-\{z,w\}}=\left|[\Phi(z)\otimes 1
  +1\otimes\Phi(w),R]\delta_w^z\right|_{w}.
\end{equation}
\item After that the commutativity of $Tr (\Phi(z)^k)$ is more or less
standard game with $r$-matrixes.
\item The functions $Tr (\Phi(z)^k)$ are invariant with respect to the action
of GL(n) by conjugation on the
space of pairs $\L(\ep),\Phi(\ep)$, so they can be pushed down to the
reduced space.
The main property of hamiltonian reduction is that it preserves the brackets
of invariant functions. So we obtain the Poisson commuting family
on the reduced space, which is as explained before the phase space
of Hitchin system.
\end{itemize}


\vskip 1cm
{\bf Acknowledgements.} The authors are grateful for their friends
and colleagues for useful and stimulating discussions:
N. Amburg, Yu. Chernyakov, V. Dolgushev, V. Kisunko, A. Kotov, D. Osipov,
 S. Shadrin, G. Sharygin, A. Zheglov, A. Zotov. A part of this work was done
during the stay of D.T. at LPTHE (Paris 6) where the principal idea for
proving lemma \ref{l1} was given by O. Babelon.

\section{Class of curves}
\subsection{Description \label{sect-desrpt-curves} }

In this section we consider the curves defined as $Spec \{ f\in \CC[z],
f(A(\epsilon))=f(B(\epsilon)) \}$, where $\ep^N=0$,
$A(\ep)=a_0+a_1\ep+a_2\ep^2+...+a_{n-1}\ep^{N-1}$ and
 $B(\ep)=b_0+b_1\ep+b_2\ep^2+...+b_{n-1}\ep^{N-1}$ are some fixed polynomial.
We start with the following obvious lemma:
{\Lem The condition
$f(A(\epsilon))=f(B(\epsilon)) $ defines a subring in $\CC[z]$.
}
\\
These
curves are singular curves with the only singular point.
If $a_0=b_0$ then the curve has the only one branch (like cusp $y^2=x^3$), if $a_0\ne b_0$
then the curve has two branches (like node $y^2=x^2(x+1)$).
More precisely we consider projective curves $\Sigma^{proj}$ which are
obtained from the curve above adding one smooth point $\infty$. (It can be
obviously formalized).
Concretely this implies the geometrical objects like differentials, endomorphisms
etc. to do not have poles at $\infty$.

{\Ex ~ }
Consider $Spec \{ f\in \CC[z],
 f(1)=f(0) \}$. This is node (or double point) curve.

{\Ex ~ }
Consider $Spec \{ f\in \CC[z],
 f(\epsilon)=f(0) \}$, where $\ep^2=0$.
One has: $f(\epsilon)=f(0)+f'(0)\epsilon$, hence
condition $f(\epsilon)=f(0)$ gives $f'(0)=0$.
So this curve is a cusp curve $y^2=x^3$.
So one can informally say that "cusp curve  is result of glueing of two
infinitely close to each other points $z=0,z=\ep$.

{\Ex ~ }
Consider $Spec \{ f\in \CC[z],
f(\epsilon)=f(0) \}$, where $\ep^N=0$.
Analogously we obtain a curve
defined by the conditions: $f'(0)=f^{\prime \prime}(0)=...=f^{(N-1)}=0$

{\Ex ~ }
 Consider $Spec \{ f\in \CC[z],
f(\epsilon)=f(\alpha\ep) \}$, where $\ep^N=0$, $\alpha$ is primitive
root of unity of order $k>1$, i.e. $\alpha^k=1$.
We obtain a curve
defined by the conditions: $f^{(l)}(0)=0$, for $l\neq k,2k,3k,...$
and $l<N$.

{\Ex ~  }
Consider $Spec \{ f\in \CC[z],
f(\epsilon)=f(-\epsilon+b_2\epsilon^2+b_3\epsilon^3) \}$, where  $\ep^4=0$.
We obtain a curve
defined by the conditions: $f'(0)=0$,
$f{'''}(0)=-3b_2 f''(0)$.
One can see that it does not depend on $b_3$.



{ \Prop
The condition $
f(A(\epsilon))=f(B(\epsilon)) $, where
$A(\ep)=a_0+a_1\ep+...$ and $a_1\ne 0,$ is equivalent to the condition
$f(a_0+\epsilon)=f(D(\epsilon)) $ for some $D(\ep)$.
}
\\
Actually, for $a_1\ne 0$ one can obviously find such polynomial $C(\ep)$ that
$A(C(\ep))=a_0+\ep$, (if $a_0=0$, then $C(\ep)$ is
truncation of inverse formal power series for $A(\ep)$)
so one sees that $D(\ep)=B(C(\ep))$.

{\Ex ~ } The condition $f(\ep+\ep^2)=f(-\ep+\ep^2)$ is equivalent
to condition $f(\ep)=f(-\ep+2\ep^2-4\ep^3)$, where $\ep^4=0$.
Here we just look for $C(\ep)$ such that $A(C(\ep))=\ep$. This  means
that $C(\ep)^2+C(\ep)=\ep$ hence $C(\ep)=\ep-\ep^2+2\ep^3$, and hence
$B(C(\ep))=-\ep+2\ep^2-4\ep^3$.

So one sees that in general position we can restrict ourselves to the
curves, where $A(\ep)=a_0+\ep$, by change of variables $z=\tilde z +a_0$
we can consider only the case $A(\ep)=\ep$.

{\Rem ~ }   The  curves  which are  given by
$Spec \{  f\in \CC[z],
f(A(\epsilon))=f(B(\epsilon)) \} $ and $Spec \{  f\in \CC[z],
f(A((\epsilon+c)))=f(B((\epsilon+c))) \}$, where $c \in \CC$ in general are
not isomorphic. For example $f(\ep)=f(0)$, $\ep^2=0$ is cusp curve,
but $f(\ep+1)=f(0)$ - is curve with two branches.


Let us calculate the genus of our singular curves.

{\Prop
\label{prop-curves}
 Here we consider three possible cases concerning the
polynomials  $A(\ep)$ and $B(\ep).$

\begin{itemize}

  \item{\bf Nilpotent:} If $A(\ep)=\ep$ and $\underbrace{B(B(B...(B}_{\mbox{k times}}
  (\ep)))=0$ $mod$
$\ep^{N-1}$, for some $k$ then
$dim
H^1({\cal O
})=N-1$ (i.e. genus equals $N-1$)
for the curve $\Sigma$ defined above. This curve is equivalent to the curve defined by $B(\ep)=0$
(i.e. the conditions defining the curve are: $f(\ep)=f(0)$ $mod$
$\ep^{N-1}$).

  \item{\bf Root of Unity:} If $A(\ep)=\ep$ and $\underbrace{B(B(B...(B}_{\mbox{k times}}
  (\ep)))=\ep$ $mod$
$\ep^{N-1}$ then
$dim H^1({\cal
O})=N-1-[(N-1)/k]$ (i.e. genus equals $N-1-[(N-1)/k]$)
for the curve $\Sigma$ defined above.
The curve depends  nontrivially  on the coefficients
$b_1, ...,b_{N-1}$, (it does not depend on $b_N$ if $b_1 \neq 1$).

If for all $k$  $\underbrace{B(B(B...(B}_{\mbox{k times}}
(\ep)))=\sum c_l\ep^l $ and all the coefficients $c_l$ are not
equal to zero  then the curve $\Sigma$ is the curve  defined by the
subring $f'(0)=f^{\prime \prime}(0)=...=f^{(N-1)}=0$. So  all such
$B(\ep)$ gives the same curve as $B(\ep)=0$.

\item{\bf Different Points:}
In the case $a_0\ne b_0$, where $A(\ep)=a_0+a_1\ep+...$,
$B(\ep)=b_0+b_1\ep+...$,
 $\ep^N=0$
the curve has two branches and $dim H^1({\cal O})=N$  (i.e. genus equals $N$).
\end{itemize}
}

%

Let us does not analyze more complicated case:
$\underbrace{B(B(B...(B}_{\mbox{k times}} (\ep)))=\ep$ $mod$ $\ep^{L}$
for some $L< N-1$.

The items 1 and 3 in proposition are quite obvious,
the proof of the item 2 will be given elsewhere.
Let us give only some motivation and example for item 2.
From the condition $f(\ep)=f(B(\ep))$ follows that
$f(\ep)=f(B(\ep))=f(B(B(\ep)))=f(B(B(B(\ep))))=...$
so it is natural to expect that if iteration process $B(B(...(B(\ep)))$ does
not stops, then one obtains infinitely many different points where
the values of polynomial $f(\ep)$ coincide so $f(\ep)=Const ~ mod ~ \ep^N$.
Let us give nontrivial example to illustrate our proposition.

{\Ex ~ } Consider the condition
$f(\ep)=f(-\ep+b_2\ep^2+b_3\ep^3+b_4\ep^4+b_5\ep^5+b_6\ep^6)$, where
$\ep^7=0$. Rewriting it explicitly one obtains: $f^{'}(0)=0$ from
coefficient at $\ep^1$, no conditions from coefficient at $\ep^2$;
$f^{3}(0)=-3b_2f^{2}(0)$ from $\ep^3$; $b_2^2=-b_3$ from $\ep^4$ (we mean
that if this condition is not satisfied then the curve will be $f^{k}(0)=0$,
$k<7$ - this case is the same as $B(\ep)=0$ and we are not interested in it now);
$f^{5}(0)=5(-2b_2f^{4}(0)+12f^{2}(0)(2b_2^3-b_4))$ from $\ep^5$;
$2b_2^4-3b_2b_4-b_5=0$ from $\ep^6$.
The main miracle is that the same conditions for $b_i$ arises
from the condition $B(B(\ep))=\ep$, where $B(\ep)=(-\ep+b_2\ep^2+b_3\ep^3+b_4\ep^4+b_5\ep^5+b_6\ep^6)$
(i.e.
$B(B(\ep))=\ep-2(b_2^2+b_3)\ep^3+b_2(b_2^2+b_3)\ep^4+
2(2b_2^4-3b_2b_4-b_5)\ep^5-3b_2(2b_2^4-3b_2b_4-b_5)\ep^6$
(in this calculation we substituted $(b_2^2+b_3)=0$, when we calculated
terms at $\ep^5, \ep^6$)).

{\Rem ~ } Also we see such a strange observation on the iteration of polynomials
$B(\ep)=-\ep+...$:
assume that $B(B(\ep))=\ep ~ mod ~ \ep^{2k+1}$ then expression for coefficients
at $\ep^{2k+2}$ and at $\ep^{2k+3}$ are proportional to each other.
Analogous situation should be with the iteration of polynomials of the type:
$B(\ep)=\alpha\ep+...$, where $\alpha$ is root of unity.

{ \Rem  ~~ } If $\underbrace{B(B(B...(B}_{\mbox{k times}} (\ep)))=\ep$ $mod$
$\ep^{N-1}$ then we can obviously choose such $b_{N-1}$
that $\underbrace{B(B(B...(B}_{\mbox{k times}} (\ep)))=\ep$ $mod$
$\ep^{N}$. The curve $\Sigma$ does not depend on $b_{N-1}$
so we do not loose generality when we consider such $B(\ep)$ that
$\underbrace{B(B(B...(B}_{\mbox{k times}} (\ep)))=\ep$ $mod$
$\ep^{N}$.

\subsection{Geometric (Schematic) Interpretation}
The curves defined as $Spec \{ f\in \CC[z], f(A)=f(B) \}$, where
$A,B \in \CC$ geometrically can be described as glueing two points
$A,B$ to each other. So one should think about our curves
$Spec \{ f\in \CC[z], f(A(\epsilon))=f(B(\epsilon)) \}$, where $\ep^N=0$,
as gluing together not two geometric points $A,B \in \CC$
but two subschemes $ Spec \{ \CC[\ep]/\ep^N \}$.
This is leading analogy in exploring the properties of these
singular curves i.e. description of dualizing sheaf, vector bundles
and their endomorphisms.
Subschemes are obviously given by $ Spec \{ \CC[\ep]/\ep^N \} \mapsto
Spec \{\CC[z] \}: z\mapsto A(\ep)$ and another subscheme $z\mapsto B(\ep)$.
For usual geometric point we can speak about "value of the function
at a point" the same can be done for the schematic point
i.e. "the value of function $f\in \CC[z]$ at a schematic point
$z\mapsto A(\ep)$" is just $f(A(\ep))$ the "value" is not a number, but
it is element of the structure ring of a point i.e. "value" belongs to
$ \{ \CC[\ep]/\ep^N \}$. So equality $  f(A(\ep))=f(B(\ep)) $
means that the "values" of function $f$ coincide in two schematic points.
So this means that we have glued this two points together.

\subsection{Further examples}

One can glue many points:
$$Spec \{ f\in \CC[z],
f(A_i(\epsilon))=f(A_j(\epsilon)), f(B_i(\epsilon))=f(B_j(\epsilon)),
f(C_i(\epsilon))=f(C_j(\epsilon)), \ldots \}$$
Most of our constructions can be generalized straightforwardly to this situation,
we will not discuss it for simplicity.

\subsection{How general is this class of singular curves ?}
Presumably not all singularities can be described as glueing subschemes.
For example the condition $f(A(\epsilon))=f(B(\epsilon))$
always implies $f^{'}(0)=0$.
Possibly all the  curves
$Spec \{ f\in \CC[z],
f(\epsilon)=f(B(\epsilon)) \}$, where
$\underbrace{B(B(B...(B}_{\mbox{k times}} (\ep))...)=\ep$ $mod$
$\ep^{N-1}$, are analytically equivalent to the curves
$Spec \{ f\in \CC[z],
f(\epsilon)=f(\alpha\epsilon) \}$, where $\alpha^k$=1.
There is another known example which does not fit into our description:
{\Ex ~} Let us consider the singular curve such that its affine part without
$\infty$ is described by the subring of $\Cz$ generated by $1,z^3,z^5.$ It
corresponds to the  Sylvester diagram $1,0,0,1,0,1,1,0.$

\section{Moduli of bundles}
\subsection{Projective Modules
\label{proj-mod}
}

In ``geometric-to-algebraic'' dictionary vector bundles correspond to
projective modules, it's well known that projective modules
are the same as locally (in Zariski topology) free modules.
Recall that starting with vector bundles we consider the sheaf of its
sections and thus obtain a sheaf of modules, which will be locally free (and so
projective), vice versa we consider sheaf of locally free modules and so we can find the glueing maps
and so define the vector bundle. We assume that reader is more or less
familiar with this geometric-to-algebraic correspondence.
Algebraic language here is more preferable.

{\Def  Let us recall that the fiber at a point $P$ of a module $M$
over a ring $R$ is defined as $M^{loc}/ I^{loc} M^{loc} $,
where $I$ is the  maximal ideal of the point $P$
and ``$loc$'' means ``localization at point $P$''.}

Recall that for locally free (projective) modules over algebras over $\CC$
the fibers at each points are of the
same dimension as $\CC$ vector spaces. We will use this fact as a test for a module to be non
projective.

{\Prop
The subset of vector-valued functions $s(z)$ on $\CC$
i.e. $s(z)\in \CC[z]^{\oplus r}$ which satisfy
condition $s(A(\ep))=\Lambda(\ep)s(B(\ep))$ for some fixed
matrix-valued polynomial $\Lambda(\ep)= \sum_{i=0...N-1}\Lambda_i\ep^i$
is a  module over  the algebra of functions
$ \{ f\in \CC[z] f(A(\epsilon))=f(B(\epsilon)) \}$.
}

We will denote this module $M_{\Lambda}$. This module is not always
projective (see next proposition). Obviously $M_{\Lambda}$ is rank $r$ for
general $\Lambda$.

{\Ex ~ \label{proj-mod-ex1} }
 Consider the double point curve:
$\Sigma= Spec \{ f\in \CC[z] : f(1)=f(0) \}$.  Then  rank $1$ modules
(line bundles and rank one torsion free sheaf) are
parameterized by the $\lambda \in \CC$. They are given by the condition
$ \{ s(z)\in \CC[z] : s(1)=\lambda s(0) \} $. Obviously $M_{\Lambda}$ are
torsion free modules. For $\lambda =0$ one can obviously see that it is not
projective module, because the fiber at the point $z=0$ jumps and becomes
two-dimensional, this is impossible for locally free modules.
It's a nice exercise to calculate the divisor of the line bundle
$\ML$. For $\lambda\neq 0$ one can obviously see that this module is locally free and hence
projective.
This example illustrates also that the moduli space of line bundles (the so-called generalized Jacobian)
on singular curve is non compact (it is $\CC^{*}$ in this case and this isomorphism is also
an isomorphism of groups, where as usually one considers the tensor product as a group operation
on line bundles). The moduli space can be
compactified by torsion free modules. In this case one should add one
module corresponding to $\lambda =0$ (it is isomorphic to the module $\lambda =\infty$, i.e.
the module $ \{ s\in \CC[z] : 0= s(0) \} $).
It can be shown that if one constructs properly
the algebraic structure on the set of torsion free sheaves of rank $1$  it
coincides with the curve $\Sigma^{proj}$ itself as a manifold.
This can be done constructing the Poincar\'e line bundle on the product of curve with
itself. This fact reflects the elliptic nature of such kind of curves.

{\Ex ~ \label{proj-mod-ex2} } Consider the cusp curve: $\Sigma= Spec \{ f\in
\CC[z] : f(\ep)=f(0) \}, \ep^2=0$. The modules can be described by $ \{ s(z)
\in \CC[z] : s(\ep)=(\lambda_0+\lambda_1\ep) s(0) \} $. Obviously if
$\lambda_0 \ne  1$ then the module $\ML$ is the zero module. So we consider
$\lambda(\ep)= 1+\lambda_1 \ep $ and the modules $\ML$ can be described
explicitly as $\{ s \in \CC[z] : s^{'}(0)=\lambda_1 s(0) \} $. It coincides
with the traditional description (\cite{CT1}, example 6).
 In this example for all $\lambda_1 \in \CC$ these modules are
projective. So $\CC$ is module space of line bundles. It can
be compactified adding one point $\lambda_1 = \infty$ (i.e. the module
$\{ s \in \CC[z] : 0= s(0) \} $, which is the same as the maximal ideal of the
singular point $z=0$, and the same as the direct image of ${\cal O}
^{norma}$ - the structure sheaf of the normalized curve,
and the same as just $\CC[z]$ considered as a module over our algebra).
Properly introduced algebraic structure will show that this
moduli space is $\Sigma ^{proj}$ itself, not $CP^1$ as one might
think from the naive point of view.

{\Ex ~ \label{proj-mod-ex3} } Consider the curve: $\Sigma= Spec \{ f\in \CC[z] : f(\ep)=f(0) \},
\ep^N=0$. Analogously one can consider modules: $ \{ s(z) \in \CC[z] :
s(\ep)=(1+\lambda_1\ep+\lambda_2\ep^2+... +\lambda_{N-1} \ep^{N-1}) s(0)
\}$. It can be rewritten as $ \{ s(z) \in \CC[z] : s^{i}(0)=\lambda_i s(0)
\}$. All of these modules will be projective. Corresponding line bundles
over the $\Sigma^{proj}$ exhaust all the line bundles of degree zero.

\noindent
{\bf Question }
It seems that there is only one torsion free module which is not
projective - i.e. maximal ideal of point $z=0$ (
the same as the direct image of ${\cal O} ^{norm}$
and the same as just $\CC[z]$ considered as a module over our algebra).
So the compactification would be some singular one point compactification
of $\CC^{N-1}$, it would be nice to understand the local ring of this
singular point.



{\Prop \label{prop-proj-mod}
As in proposition \ref{prop-curves} we treat three cases:

\begin{itemize}
  \item {\bf Nilpotent case:} $M_{\Lambda}$ is projective module over the algebra $ \{
f\in \CC[z] : f(\epsilon)=f(B(\epsilon)) \}$, where
$\underbrace{B(B(B...(B}_{\mbox{k times}} (\ep)))=0$; $\ep^{N}=0$
{\bf iff } $\Lambda(0)=\Lambda_0=Id.$
\item {\bf Root of unity case:} $M_{\Lambda}$ is projective module over the algebra $ \{
f\in \CC[z] : f(\epsilon)=f(B(\epsilon)) \}$, where
$\underbrace{B(B(B...(B}_{\mbox{k times}} (\ep)))=\ep$;
$\ep^{N}=0$
\\ %
 ~ { \bf iff }
$\Lambda(\ep)\Lambda(B(\ep))\Lambda(B(B(\ep)))...\Lambda
(\underbrace{B(B(B...(B}_{\mbox{k-1 times}} (\ep))))=Id$, where
$Id$ is identity matrix (it does not depend on $\ep$).
\item  {\bf Different Points
} $M_{\Lambda}$
is projective module  over the
algebra $ \{ f\in \CC[z]: f(A(\epsilon))=f(B(\epsilon)) \}$, where $a_0\neq
b_0$ {\bf  iff} $\Lambda_0$ is invertible matrix.
\end{itemize}
 }

The proof will be given elsewhere. The items 1 and 3 are obvious.
Let us illustrate the item 2.

{\Ex ~ } Consider the curve: $\Sigma= Spec \{ f\in \CC[z] : f(\ep)=f(-\ep) \},
\ep^4=0$. So here $B(B(\ep))=\ep$. Explicitly it can be described as
$\Sigma= Spec \{ 1, z^2, z^4,z^5,... \}$.
One can consider modules: $ \{ s(z) \in \CC[z] :
s(\ep)=(1+\lambda_1\ep+\lambda_2\ep^2+\lambda_{3} \ep^{3}) s(-\ep)
\}$. From proposition \ref{prop-curves}  we know that    $dim H^1({\cal O})=2$,
it's well-known that $dim H^1({\cal O})$ is tangent space to moduli space
of line bundles, so moduli space of line bundles on our curve
should be two-dimensional, on the other hand we see three parameter
$ \lambda_1, \lambda_2, \lambda_{3}$ which describes the rank $1$ modules
(which are obviously torsion free), so in order to solve the contradiction
we should show that not all of modules are projective.
Let us rewrite condition $ s(\ep)=(1+\lambda_1\ep+\lambda_2\ep^2+\lambda_{3} \ep^{3}) s(-\ep)
$ explicitly. The coefficient at $\ep^1$ will give:
$s^{'}(0)=\lambda_1 s(0)/2$;
the coefficient at $\ep^2$ will give:
$\lambda_1^2/2 s(0)=\lambda_2 s(0) $, we will show that
if $\lambda_2\ne\lambda_1^2/2$ then the
module is not projective.
The coefficient at $\ep^3$ gives:
$s^{(3)}(0)=3 (\lambda_3 s(0)+\lambda_1 s^{(2)}(0)/2-\lambda_1^3/4 s(0))$
(we used $\lambda_1^2/2 s(0)=\lambda_2 s(0) $ in the last term).
So we see that if the condition $\lambda_1^2/2=\lambda_2  $
is not true then the module  can be described as $s(0)=0, s'(0)=0, s^{(3)}(0)=3/2 \lambda_1
s^{(2)}(0)$. Hence it's fiber at point zero is two-dimensional,
(in general point it is $1$-dimensional) so this module cannot be locally free
(=projective).

We obtained the condition $\lambda_2=\lambda_1^2/2.$ It is rather miraculous
but simple to demonstrate
that this condition is equivalent to the one stated above: $\Lambda(\ep)\Lambda(B(\ep))=Id.$
On the other hand there is a following motivation:
we request $s(\ep)=\Lambda(\ep)s(B(\ep))$, so
\begin{equation*}
\begin{split}
s(\ep)&=\Lambda(\ep)s(B(\ep))=
\Lambda(\ep)\Lambda(B(\ep))s(B(B(\ep)))= \ldots \\
&= \Lambda(\ep)\Lambda(B(\ep))\Lambda(B(B(\ep))) \ldots \Lambda
(\underbrace{B(B(B\ldots(B}_{\mbox{k-1 times}} (\ep)))) s((\underbrace{B(B(B\ldots(B}_{\mbox{k times}}
(\ep)))) \\
&= \Lambda(\ep)\Lambda(B(\ep))\Lambda(B(B(\ep)))\ldots\Lambda
(\underbrace{B(B(B\ldots(B}_{\mbox{k-1 times}} (\ep)))) s(\ep).
\end{split}
\end{equation*}
The condition $\underbrace{B(B(B...(B}_{\mbox{k times}} (\ep)))=\ep$ makes it natural to expect
that $$\Lambda(\ep)\Lambda(B(\ep))\Lambda(B(B(\ep)))...\Lambda
(\underbrace{B(B(B...(B}_{\mbox{k-1 times}} (\ep))))=Id,$$ otherwise
as we have seen in the example the fiber at zero jumps and the module
is not projective, this is true in general.


\subsection{Vector Bundles \label{sect-vect-bund}}

The modules $M_\Lambda$ are equivalent, if there exists
invertible map of modules $K(z):M_\Lambda\mapsto M_{\tilde\Lambda}$.

Recall that we denoted by $\Sigma^{proj}$ the projective curve
which we obtain from the affine  curve
$Spec \{ f\in \CC[z] f(\epsilon)=f(B(\epsilon)) \}$ by adding one smooth
point at infinity.  The module $M_{\Lambda}$ gives a vector bundle over
$\Sigma^{proj}$ in an obvious way: we define the sheaf
which is trivial rank $r$ module over the chart containing infinity
and not containing singular point and which is module $\ML$ (
or more precisely its localization) over the chart which contains the
singular point. The degree of such bundles equals zero.
The vector bundles are equivalent if there exists
the invertible map of modules $K(z):M_\Lambda \mapsto M_{\tilde\Lambda}$
over each chart. So we see that $K(z)$ is matrix polynomial which
should be regular both at infinity and zero so $K(z)$ does not depend on $z$
so it is constant.
 In this way we obtain:

{\Prop
The vector bundles  $\ML$ over $\Sigma^{proj}$  are isomorphic if there
exists a constant matrix $K$ such that
$\Lambda(\ep)=K {\tilde\Lambda(\ep)}K^{-1} $.
}

As a corollary we obtain the following:
{\Th \label{bund-on-sigm}
The open subset in the  moduli space of vector bundles of degree zero
and rank $r$ over the curve $\Sigma^{proj}$
is the space of matrix polynomials
$\Lambda(\ep)=\sum_{i=0...N-1}\Lambda_i \ep^i$, which satisfy the
conditions below, factorized by the conjugation by constant
matrices. The conditions for $\Lambda(\ep)$ are the following:
\begin{itemize}
  \item {\bf Nilpotent case:} for
the curves
$
Spec \{ f\in \CC[z]: f(\epsilon)=f(B(\epsilon)) \}$, where
$\underbrace{B(B(B...(B}_{\mbox{k times}} (\ep)))=0$ $mod$
$\ep^{N-1}$ \\ we request $\Lambda_0=Id$.
  \item {\bf Root of Unity case:} for
the curves
$
Spec \{ f\in \CC[z]: f(\epsilon)=f(B(\epsilon)) \}$, where
$\underbrace{B(B(B...(B}_{\mbox{k times}} (\ep)))=\ep$\\
we request
$\Lambda(\ep)\Lambda(B(\ep))\Lambda(B(B(\ep)))...\Lambda
(\underbrace{B(B(B...(B}_{\mbox{k-1 times}} (\ep))))=Id$ and
$\Lambda_0=Id$.
  \item {\bf Different points:} For the curves $ Spec \{ f\in \CC[z]:
f(A(\epsilon))=f(B(\epsilon)) \}$, where $a_0\neq b_0$ we request
$\Lambda_0$ to be invertible.
  \end{itemize}
  }

%

{\Rem ~} Let us omit the questions of stability of bundles and what exactly we
mean by "factor".  Let us also note that all the bundles $\ML$ satisfy the
property that $\pi^* \ML$ will be trivial bundle on $\CC P^1$, where $\pi: \CC
P^1 \to \Sigma^{proj}$ is normalization map.  Obviously there are lots of
bundles $\F$ of degree zero on
$\Sigma^{proj}$ such that $\pi^* \F$ are not trivial
bundles but some bundles of the type $\oplus_{k=1,...r} \O (t_k)$, such that
$\sum t_k=0$.  So by no means we
do not obtain all bundles on $\Sigma^{proj}$ as
bundles $\ML$ for some $\L$.  But nevertheless general stable and possibly
semistable bundle satisfy the property that $\pi^* \F $ will be trivial
bundle on $\CC P^1$, and it is easy to see from our previous description of
projective modules that general stable bundles can be obtained as bundles
$\ML$ for some $\L$.

\subsection{Geometric interpretation}

Geometrically the curve
$Spec \{ f\in \CC[z]:
f(A(\epsilon))=f(B(\epsilon)) \}$ is obtained by glueing
two points $A$ and $B$ together.
We described the modules over this curve as
$s(z): s(A(\ep))=\Lambda(\ep)s(B(\ep))$. Geometrically this
can be interpreted as glueing the fibers at point $A$ and $B$
 of the trivial bundle by the map $\Lambda:$ fiber at $A\mapsto$
fiber at $B$. Indeed, fiber of the module M at point $A$
is $M_{A}=M\otimes_{{\cal O}(\Sigma)} {{\cal O}_{(A)}}$
in our case $M_B=M_A= \CC[\ep] / \ep^N$ so
we need to describe the  map $\Lambda:\CC[\ep] / \ep^N \mapsto
\CC[\ep] / \ep^N$  which is the map of modules over algebra $\CC[\ep] /
\ep^N$ it is given by it's value at $1$ which
we denote by $\Lambda(\ep)$, so the map
$\Lambda:\CC[\ep] / \ep^N \mapsto
\CC[\ep] / \ep^N$ is given by the multiplication on $\Lambda(\ep)$.
The condition for $s(z): s(A(\ep))=\Lambda(\ep)s(B(\ep))$
geometrically means that we consider such $s(z)$ that
its value in point $A$ equals to $\Lambda$ multiplied its value at point $B$
for the map $\Lambda:M_{B}\mapsto M_{A}$ described above.

\subsection{On the notation $\frac{1}{\ep}$ for $\ep^N=0$}

It's obvious that if $\ep^N=0$ then  one cannot
introduce the inverse element $\frac{1}{\ep}$
in a sense of the associative multiplication,
(for example: let $\ep^2=0$  $ ((\frac{1}{\ep} \ep ) \ep)
=(1 ) \ep = \ep \ne  (\frac{1}{\ep} ( \ep  \ep))
= (\frac{1}{\ep} ( 0 ))=0 $).
Nevertheless as we will see from the lemma below
there is no problem in using formal expressions with negative degrees in
$\ep$ to define the paring.
It will be very convenient for us to use such notation
in next sections where we will describe differentials
on  singular curves.
\\
{\bf Notation } Let us settle that
$ Res_{\ep} A = A_{-1} $, where $A=\sum_i  A_i \ep^i$.

{ \Lem Let us consider $f(\ep)
=\sum_{i \ge -N} f_i \ep^i $,  $g({\ep}) =\sum_{i\ge 0} g_i \ep^i$ ,
 $h(\ep) =\sum_{i\ge 0} h_i \ep^i$  then
\bea
Res_{\ep}  ((g h)f) = Res_{\ep} (g(h f))
\eea
}
\\
The proof is obvious.

In the text below we will use terms containing $\frac{1}{\ep}$
only in expressions like in the lemma above, so we do not need
to care about the way of putting the brackets.

Let us make some remarks.
What should be considered as a module of meromorphic differentials
on the scheme $Spec \{ \CC[\ep]: \ep^N=0 \}$ ?
It seems to be natural to denote by such
differentials the expressions
$f(\frac{1}{\ep}) d\ep
=\sum_{i =-N,..., -1 } f_i \ep^i d\ep $.
The structure of module is given by the
usual multiplication combined with the dropping out
all the terms $\ep^k$ where $k$ is out of range
$k = -N,..., -1 $.
As a module it is just trivial module. The
use of such notations for the elements of the trivial
module is convenient, because the pairing
between the differentials $\omega$ and functions $f$
can be written in the same way as usually:
$ Res_{\ep} f \omega$.

{ \Lem Consider $f(z,\ep)=\sum_{i,j} f_{i,j} \ep^iz^j $, it is
obviously true that:
\bea
Res_{\ep} Res_z  f =  Res_z  Res_{\ep} f = f_{-1,-1}
\eea
}

{ \Lem Consider $f(z)=\sum_{i\ge 0 } f_{i} z^i $
and $A(\ep)=\sum_i a_i\ep^i$, where $\ep^N=0$,
 it is obviously true that:
\bea
Res_{z} \frac{  f(z)}{z-A(\ep)} =  f(A(\ep))
\eea
}

\section{Dualizing Sheaf}
\subsection{Construction}

Recall that the dualizing sheaf $\mathcal{K}$ on manifold $X^{n}$ is such sheaf that
there is isomorphism $t:$ $H^{n}(X,\mathcal{K})=\CC$ and for any other
coherent sheaf $F$ there is natural pairing
$Hom(F,\mathcal{K})\times H^n(X,F)\to H^n(X,\mathcal{K})$, which is combined with
isomorphism $t$ gives isomorphism $Hom(F,\mathcal{K})\to H^n(X,F)^{*}$,
where $V^*$ is space dual to $V$.
So for the flat sheaves $F$ we have isomorphism
$H^0(K\otimes F^{*})\to H^{n}(F)^{*}$.

On a nonsingular curve the dualizing sheaf is the sheaf of holomorphic $1$-forms,
but for a singular curve, one should specify what are $1-forms$ and the naive definition
i.e.  the Kaehler differentials (\cite{Harts}
ch. 2 sect. 8) is not the right object. It's known that
(see \cite{Serre,HM}):

{\Prop The dualizing sheaf $\mathcal{K}$ on the singular curve $\Sigma$ can be described
as: sheaf of meromorphic $1$-forms $\alpha$  on
normalization,
such that $\forall f \forall P\in \Sigma$ it is true that:
\bea \label{hol-dif}
\sum_{P_i \in
 \Sigma^{norm}: \phi(P_i)=P} Res_{P_i} f\alpha =0,
\eea
 where $f$
 is the pullback of a function
 $f$ on the singular curve to its normalization, $P_i$ are such points on
the normalization that they maps to the same point $P$ on the singular curve
 under the normalization map $\phi: \Sigma^{norm}\mapsto \Sigma$.

 }

 We will call the elements of $H^0(\mathcal{K})$ as holomorphic differentials
 on curve $\Sigma$.

{\Rem ~ } The description of holomorphic differentials
given above is very natural for the following reason:
consider the nonsingular curve, then obviously $Res_{p} f w  =0$
for any point $p$, holomorphic function $f$ and fixed holomorphic
differential $w$, in singular situation we see that there are some
meromorphic differentials on normalization which satisfy
analogous property that $Res_{p} f w  =0$ for any $f$ which comes
from singular curve, so it is natural to guess that they should
be treated as holomorphic. The more rigorous argument consists in
demonstrating that this realization of the canonical sheaf really provide
the duality $$H^k(F\otimes \mathcal{K})\times H^{n-k}(F^*)\longrightarrow
\C.$$ For some examples and demonstration see \cite{CT1}.

{\Rem ~ } For all our curves  $ Spec \{ f\in \CC[z]:
f(A(\epsilon))=f(B(\epsilon)) \}$ the normalization is the complex plane $\CC$.
So holomorphic differentials on such singular curves are
meromorphic differentials on $\CC$ which satisfy the condition \ref{hol-dif}.


{\Ex ~ \label{h-d-ex1} }
Consider the node (or double point) curve $Spec \{ f\in \CC[z],
 f(1)=f(0) \}$. The holomorphic differentials on such curve
are meromorphic differentials on $\CC$ (which is a normalization of the node)
of the kind $w(z)= c(\frac{1}{z-1}- \frac{1}{z})dz + f(z)dz $,
where $c$ is a constant,  $f(z)$ is holomorphic function on $\CC$.
One easily sees that the condition \ref{hol-dif}, which in this case
says that the residues at points $z=1$ and $z=0$ are of the different sign, is satisfied.
So global holomorphic differentials on the corresponding  projective
curve will be of the form $c(\frac{1}{z-1}- \frac{1}{z})dz  $

{\Ex ~ \label{h-d-ex2} }
Consider the cusp curve $Spec \{ f\in \CC[z],
 f(\epsilon)=f(0) \}$, where $\ep^2=0$.
Analogously to the previous example $w(z)= \frac{cdz}{z^2} + f(z)dz $.
So global holomorphic differentials on the corresponding  projective
curve will be of the form $\frac{cdz}{z^2}$.

{\Rem ~} In two preceding examples the global holomorphic differentials can
be obtained by degenerating the elliptic nonsingular curve and the
holomorphic differentials on it (see \cite{CT1} for comments).

{\Ex ~ \label{h-d-ex3} }
Consider $Spec \{ f\in \CC[z],
f(\epsilon)=f(0) \}$, where $\ep^N=0$.
Analogously to the previous example $w(z)= \sum_{i=1}^{N-1}\frac{c_idz}{z^{i+1}} + f(z)dz $.
Hence global holomorphic differentials on the corresponding  projective
curve will be of the form $\sum_{i=1}^{N-1}\frac{c_idz}{z^{i+1}}$.

{\Ex ~  \label{h-d-ex4} }
 Consider $Spec \{ f\in \CC[z],
f(\epsilon)=f(\alpha\ep) \}$, where $\ep^N=0$, $\alpha$ is primitive
root of unity of order $k>1$, i.e. $\alpha^k=1$.
Holomorphic differentials are given by
 $w(z)= \sum_{i=1}^{N-1}\frac{c_idz}{z^{i+1}} +f(z)dz $, where $\forall l$
$c_{lk+1}=0$.

{\Ex ~ \label{h-d-ex5}}
Analogously for the curve $Spec \mathbb{C}[1, z^{k_1},
z^{k_2},...,z^{k_n}]$ we see that holomorphic differentials will be:
all $\frac{c_p dz}{z^{p+1}}$, such that $p\ne k_i$.

{\Ex ~  \label{h-d-ex6} }
Consider $Spec \{ f\in \CC[z],
f(\epsilon)=f(-\epsilon+b_2\epsilon^2+b_3\epsilon^3) \}$, where  $\ep^4=0$.
Holomorphic differentials are given by
 $w(z)=\frac{c_1dz}{z^{2}} +c_2(\frac{1}{z^{4}}+\frac{b_2}{z^{3}}) dz+f(z)dz
$.

{\Ex ~  \label{h-d-ex7} }
Consider $Spec \{ f\in \CC[z],
f(\alpha \epsilon)=f(1-\beta \epsilon ) \}$, where  $\ep^2=0$.
Holomorphic differentials are given by
 $w(z)=c_1(\frac{1}{z}-\frac{1}{z-1}) dz
 +c_2(\frac{\alpha}{z^{2}}+\frac{\beta}{(z-1)^{2}}) dz+f(z)dz
$.

\subsection{Global sections
\label{sect-dual-expl}
}

In the proposition above the holomorphic differentials (i.e. the
sections of dualizing sheaf)  were given
not explicitly, and to find them one should solve a system of linear
equations.
The key proposition (see below)  of this section shows that for our curves
$\Sigma =Spec \{ f\in
\CC[z],f(A(\epsilon))=f(B(\epsilon)) \}$   the differentials can be given
explicitly, moreover we will see later that the symplectic form on the
cotangent bundle to the moduli space
in such coordinates will be written explicitly.


 {\Prop
\label{hol-dif-main}
 For all $\omega_i \in \CC, i=0,...,N-1$ and their generating function
$\omega(\ep)=\sum_{i=0...N-1 } \omega_i\frac{1}{\ep^{i+1}}$
the expression
\bea \label{tilde-phi}
\tilde \omega (z)=
Res_{\ep} \frac{\omega(\ep)dz}{z-A(\ep)} - \frac{\omega(\ep)dz}{z-B(\ep)},
\eea
gives a holomorphic differential (i.e. an elements of $H^0(\K)$) on the
curve $\Sigma^{proj}$,
defined on the affine chart without
infinity as $ Spec \{ f\in \CC[z],f(A(\epsilon))=f(B(\epsilon)) \}$.
Moreover any holomorphic differential can be obtained in this way.
}

The correspondence $\omega_i \mapsto \tilde \omega (z)$ is not in general injective
(see examples below).

 Expression $\frac{1}{z-A(\ep)}$ should be understood
expanding the geometric progression in the region $z-a_0>>\ep$.

 It is easy to see that for any $f\in \CC[z]$, which is a function on our curve,
i.e. satisfies the conditions $f(A(\epsilon))=f(B(\epsilon))$,
it is true  that:
 $Res_z f Res_{\ep}(
 \frac{\omega(\ep)dz}{z-A(\ep)} - \frac{\omega(\ep)dz}{z-B(\ep)}) =0,$
 so $\tilde \omega (z)$  is really a holomorphic differential on our
 singular curve. It can be shown that our construction gives
 all elements in $H^0(\K)$.

 Thus the coefficients $\omega_i$ parameterize the space $H^0(K)$.
In general they can be dependent, but it's easy to choose independent
parameters. We will see that in terms of $\omega_i$ the symplectic form
 on the cotangent bundle to the moduli space
of vector bundles on $\Sigma$ can be written explicitly and quite simply.

{\Ex ~ }
Consider the node (or double point) curve $Spec \{ f\in \CC[z],
 f(1)=f(0) \}$. We take $\omega(\ep)=\omega_0\frac{1}{\ep}$.
The formula \ref{tilde-phi} gives:
$\tilde \omega(z)=
Res_{\ep} \frac{\omega_0 dz}{\ep(z-1)} - \frac{\omega_0dz}{\ep z}=
\omega_0 ( \frac{dz}{z-1} - \frac{dz}{z})$.
It coincides with the holomorphic differential from example
\ref{h-d-ex1}.

{\Ex ~ }
Consider the curve  $Spec \{ f\in \CC[z],
 f(\epsilon)=f(0) \}$, where $\ep^N=0$.
We take $\omega(\ep)=\omega_0\frac{1}{\ep} + \omega_1\frac{1}{\ep^2}+...+ \omega_{N-1}\frac{1}{\ep^{N}} $.
The formula \ref{tilde-phi} gives:
\begin{equation*}
\begin{split}
\tilde \omega(z)&=
Res_{\ep}
\omega(\ep)
( \frac{dz}{z-\ep} - \frac{dz}{z})
= Res_{\ep}
\omega(\ep)
 (\frac{dz}{ z(1-\frac{\ep}{z})} - \frac{dz}{ z})\\
&= Res_{\ep}
\omega(\ep)
( \frac{dz}{z}(1+\frac{\ep}{z}+(\frac{\ep}{z})^2+...+(\frac{\ep}{z})^{N-1} ) -
\frac{dz}{z})\\
&= Res_{\ep}
(\omega_0\frac{1}{\ep} + \omega_1\frac{1}{\ep^2} +...+ \omega_{N-1}\frac{1}{\ep^{N}} )
(\sum_{i=1,...,N-1} \frac{\ep^{i}dz}{z^{i+1}} )
=\sum_{i=1,...,N-1} \frac{ \omega_idz }{z^{i+1}}.
\end{split}
\end{equation*}
It coincides with the holomorphic differentials from example
\ref{h-d-ex3}.

{\Ex ~ }
 Consider $Spec \{ f\in \CC[z],
f(\epsilon)=f(\alpha\ep) \}$, where $\ep^N=0$, $\alpha$ is a primitive
root of unity of order $k>1$, i.e. $\alpha^k=1$.
We take $\omega(\ep)=\omega_0\frac{1}{\ep} + \omega_1\frac{1}{\ep^2}+...+ \omega_{N-1}\frac{1}{\ep^{N}} $.
The formula \ref{tilde-phi} gives:
$$\tilde \omega(z)=
Res_{\ep}
\omega(\ep)
( \frac{dz}{z-\ep} - \frac{dz}{z-\alpha\ep })
=\sum_{i=1,...,N-1; i\neq k, 2k, 3k , ....} \frac{ \omega_idz }{z^{i+1}}.
$$
It coincides with the holomorphic differentials from example
\ref{h-d-ex4}. And we see that the expression for holomorphic differentials
does not depend on $\omega_i$, where $ i\neq k, 2k, 3k $. So the map
$ \omega(\ep)\mapsto \tilde \omega(z)$ is not injective.

{\Ex ~  }
Consider $Spec \{ f\in \CC[z],
f(\epsilon)=f(-\epsilon+b_2\epsilon^2+b_3\epsilon^3) \}$, where  $\ep^4=0$.
We take $$\omega(\ep)=\omega_0\frac{1}{\ep} + \omega_1\frac{1}{\ep^2}+
\omega_2\frac{1}{\ep^3}+ \omega_{3}\frac{1}{\ep^{4}}. $$
The formula \ref{tilde-phi} gives:
\begin{equation*}
\begin{split}
\tilde \omega(z)&=
Res_{\ep}
(\omega_0\frac{1}{\ep} + \omega_1\frac{1}{\ep^2}+\omega_2\frac{1}{\ep^3}+ \omega_{3}\frac{1}{\ep^{4}} )
( \frac{2\ep dz}{z^2} +  \ep^2( -\frac{b_2}{z^2}) dz
+\ep^3( \frac{
1}{z^4}-\frac{b_3}{z^2}+\frac{2b_2}{z^3}+\frac{1}{z^4}) dz)\\
&=(\omega_1-\omega_2 b_2-\omega_3 b_3)\frac{dz}{z^2}+\omega_3(\frac{2}{z^4}+\frac{2b_2}{z^3})dz
.
\end{split}
\end{equation*}
It coincides with the holomorphic differentials from example
\ref{h-d-ex6}. And we see that the expression for holomorphic differentials
depends only on $\omega_3$ and the linear combination
$(\omega_1-\omega_2 b_2-\omega_3 b_3)$.

\subsection{Geometric interpretation ?}

In previous section we saw the geometric sense
of our constructions: curves were defined by glueing subschemes
and modules were defined by glueing fibers of trivial modules.
What is the geometric sense of our description
of differentials (proposition \ref{hol-dif-main}) ?
At the moment we do not know.

Let us speculate a little about the proposition \ref{hol-dif-main}, which
we consider as rather nice. When one glues two ordinary points
$\Sigma =Spec \{ f\in
\CC[z],f(A)=f(B) \}$ we see that holomorphic differentials are given
by $(\frac{1}{z-A}- \frac{1}{z-B})dz.$ The main property of this differential is that
its resides at two points $A,B$ coincides up to sign.
It would be tempting to do something similar in the case, when
we glue two subschemes $A(\ep), B(\ep)$, i.e. it would be nice to
introduce the notion of reside in schematic point and
to have the following proposition: holomorphic differentials
are those, whose resides in the schematic points coincide up to sign.
But as far as we were able to learn there is no such notion of residue
and our naive attempts to use for example Poincare formula, shows
that it is impossible to have such proposition. But we suspect that the formula
introduced ad hoc
$$Res_{\ep} \frac{\omega(\ep)dz}{z-A(\ep)} - \frac{\omega(\ep)dz}{z-B(\ep)}$$
should have some geometric sense and should be an example of some general
principle.

\section{Endomorphisms of $\ML$}
\subsection{Description}
{\Prop \label{end-prop}
Endomorphisms of the module $\ML$ can be described as matrix polynomials
$\Phi(z): s(z)\mapsto \Phi(z)s(z)$, which satisfy the condition
\bea\Phi(A(\ep))=\Lambda(\ep)\Phi(B(\ep))\Lambda(\ep)^{-1}.
\eea
}
\\
{\bf Proof~ } The condition above is obviously the condition for
$\Phi(z)s(z)$ to satisfy the
condition $\Phi(A(\ep))s(A(\ep))=\Lambda(\ep)\Phi(B(\ep))s(B(\ep))$,
so $\Phi(z)s(z)$ is again an element of $\ML$ and
$\Phi(z): s(z)\mapsto \Phi(z)s(z)$ is endomorphism of $\ML$.

{\Ex ~  \label{end-ex1} } In abelian situation
(i.e. rank $1$ modules over any manifold) the condition above is
empty and any element $\Phi(z)$ defines an endomorphism i.e. sheaf of
endomorphisms of any rank $1$ coherent sheaf is just $\cal O$.

{\Ex ~  \label{end-ex2} }
Consider the node (or double point) curve $Spec \{ f\in \CC[z],
 f(1)=f(0) \}$. The endomorphisms of the module $\ML$ (which is defined
as $ \{ s\in \CC[z]^r,  s(1)=\Lambda s(0) \}$, for some matrix $\Lambda$,
(see section \ref{proj-mod}, example \ref{proj-mod-ex1} )
are given by such matrix-valued polynomials
$\Phi(z)=\Phi_0+\Phi_1 z +\Phi_2 z^2+...$ which satisfy
$\Phi(1)= \Lambda \Phi(0)\Lambda^{-1} $.
Hence $\Phi (z)= \Phi_0 + (\Lambda \Phi_0 \Lambda^{-1} - \Phi_0)z
+z(z-1)\tilde\Phi(z)$, where $\tilde \Phi(z)$ is arbitrary.
When one considers the projectivization of our curve and bundle
corresponding to $\ML$ on it, one should only consider the constant
endomorphism $\Phi(z)=\Phi_0$ in order to be regular at infinity.
The condition $\Phi(1)= \Lambda \Phi(0)\Lambda^{-1} $ is satisfied if
the matrices $\Phi_0$ commute with $\Lambda$.
As a corollary we see that for general
$\Lambda $ there are only $r$-dimensional space of such matrices.
For the trivial bundle any matrix $\Phi_0$ gives its endomorphism.

{\Ex ~  \label{end-ex21} }
Consider the node (or double point) curve $Spec \{ f\in \CC[z],
 f(A)=f(B) \}$. The endomorphisms of the module $\ML$
are given by such matrix-valued polynomials
$\Phi(z)=\Phi_0+\Phi_1 z +\Phi_2 z^2+...$ that
$\Phi(A)= \Lambda \Phi(B)\Lambda^{-1} $.
Hence $\Phi (z)= \Phi_0 + \Phi_1 z
+(z-A)(z-B)(\tilde \Phi(z)) $,
where $\Phi_0, \Phi_1 $ should satisfy
$\Phi_0 +A \Phi_1= \Lambda ( \Phi_0 + B \Phi_1 )\Lambda^{-1} $ and
$\tilde \Phi(z)$ is arbitrary.
It's impossible to express in a simple way $\Phi_0$ via $\Phi_1$,
or vise versa. But it's possible to express both of them throw
the one arbitrary matrix $\Theta$.
The most simple way to see it is the following: let us write
$ \Phi (z)= \Theta_1 (z-A)-\Theta_2 (z-B)$
so the condition
$\Phi(A)= \Lambda \Phi(B)\Lambda^{-1} $
gives $\Theta_2= \Lambda \Theta_1 \Lambda^{-1} $.
$\Theta_1$ can be taken arbitrary and one finds
$\Phi(z)= \Theta_1 (z-A) - \Lambda \Theta_1 \Lambda^{-1} (z-B) $.
Hence global endomorphisms are given by  $\Phi(z)= \Theta_1 (B-A)$,
with such $\Theta_1$, which commute with $\Lambda$.

{\bf Question } We saw in example above
 that in the case of glueing of
ordinary points $f(A)=f(B)$ we have an explicit
parameterization for endomorphisms of $\ML$
i.e. all solutions of equation
$\Phi(A)= \Lambda \Phi(B)\Lambda^{-1} $
are given by
$\Phi (z)= \Theta_1 (z-A) - \Lambda \Theta_1 \Lambda^{-1} (z-B)
+(z-A)(z-B)(\tilde \Phi(z)) $
with $ \Theta_1, \tilde \Phi(z)$ being arbitrary.
The question: is it possible to give
parameterization of endomorphisms in more general case
of the curves given by glueing subschemes: $f(A(\ep))=f(B(\ep))$,
i.e. to solve the equation
$\Phi(A(\ep))= \Lambda(\ep) \Phi(B(\ep))\Lambda^{-1}(\ep)$?
In the next section we give the explicit parameterization
of the sections of $End (\ML)\otimes \K$.
But strangely at the moment we do not know how to parameterize
$End(\ML)$ itself.

{\Ex ~ \label{end-ex3} }
Consider the cusp curve  $Spec \{ f\in \CC[z],
 f(\epsilon)=f(0) \}$, where $\ep^2=0$.
So the endomorphisms of the module $\ML$ (which is defined
as $ \{ s\in \CC[z]^r,  s(\ep)=(Id+\Lambda_1\ep) s(0) \}$, for some matrix $\Lambda_1$,
see section \ref{proj-mod}, example \ref{proj-mod-ex2} )
are given by such matrix-valued polynomials $\Phi(z)=\Phi_0+\Phi_1 z +\Phi_2 z^2+...$ that
$\Phi(\ep)= \Lambda(\ep) \Phi(0)\Lambda^{-1} (\ep) $.
This is equivalent to the condition $\Phi_1=[\Lambda_1,\Phi_0]$.
When one considers the projectivization of our curve and the bundle
corresponding to $\ML$ on it,
one should consider only constant endomorphisms $\Phi(z)=\Phi_0$.
Thus one have $\Phi_1=0$ and $[\Lambda_1,\Phi_0]=0$.
Endomorphisms of the bundle $\ML$ are given by constant
matrix polynomials $\Phi(z)=\Phi_0$, where
matrices $\Phi_0$ commute with $\Lambda_1$.
As a corollary we see that for general
$\Lambda_1 $ there is only $r$-dimensional space of such matrices
(general bundles are semistable, but not stable). For the
trivial bundle any matrix gives an endomorphism.

{\Ex ~ \label{end-ex4} }
 Consider $Spec \{ f\in \CC[z],
f(\epsilon)=f(0) \}$, where $\ep^3=0$.
The endomorphisms of the module $\ML$ (which is defined
as $ \{ s\in \CC[z]^r,  s(\ep)=(Id+\Lambda_1\ep+\Lambda_2\ep^2) s(0) \}$,
for some matrices $\Lambda_1, \Lambda_2$,
see section \ref{proj-mod}, example \ref{proj-mod-ex3} )
are given by such matrix-valued polynomials $\Phi(z)=\Phi_0+\Phi_1 z +\Phi_2 z^2+...$, that
$\Phi(\ep)= \Lambda(\ep) \Phi(0)\Lambda^{-1} (\ep) $.
This is equivalent to the conditions: $\Phi_1=[\Lambda_1,\Phi_0]$,
$\Phi_2 = [\Lambda_2,\Phi_0]+[\Phi_0,\Lambda_1]\Lambda_1$.
On the projectivization of our curve one
should consider only constant endomorphism $\Phi(z)=\Phi_0$ of the bundle
corresponding to $\ML$.
One have $\Phi_1=0, \Phi_2=0$, hence $[\Lambda_1,\Phi_0]=0,
[\Lambda_2,\Phi_0]+[\Phi_0,\Lambda_1]\Lambda_1=0$.
We have two conditions for one matrix $\Phi_0$.
For general
$\Lambda_i $ only scalar matrixes can satisfy both conditions, thus
for general bundle $H^{0}(End(\ML))=\CC$
(general bundles are stable).

{\Rem ~ } In our simple examples we see that
for Calabi-Yau manifolds (the cusp curve in our case) general bundles are semistable, but
not stable; for general type manifolds (canonical sheaf is ample)
(\ref{end-ex4} in our case, which is of genus $2$) the general bundles
are stable.

\subsection{$End(\ML)\otimes \K$ \label{end-tens-k}}

{\Prop \label{hol-dif-k}
Sections of $End(\ML)\otimes \K$ over
the chart without infinity can be described
as matrix polynomials:
\bea
\label{hol-dif-k-formula}
\Phi (z)=
(Res_{\ep} \frac{\Phi(\ep)dz}{z-A(\ep)} -
\frac{\Lambda^{-1}(\ep)\Phi(\ep)\Lambda(\ep)dz}{z-B(\ep)}) +
\mbox{holomorphic in z terms},
\eea
where $\Phi(\ep)=\sum_{i=0...N-1 }
\Phi_i\frac{1}{\ep^{i+1}}$ is the formal generating function for matrices $
\Phi_i$. So we say that taking an arbitrary $\Phi_i$ and considering
$\Phi(z)$ given by the formula above, one obtains that $\Phi(z)$ is an
element of $End(\ML)\otimes
\K$.
The global sections (regular everywhere including infinity) can be obtained
 requesting additional condition for $\Phi(\ep)$:
 \bea \label{hol-dif-k-cond}
Res_{\ep} \Phi(\ep) -\Lambda^{-1}(\ep)\Phi(\ep)\Lambda(\ep)=0.
\eea
}

The proof will be given elsewhere.
The part of the proposition about the global sections follows
trivially from the first part, because the condition \ref{hol-dif-k-cond}
is just the condition for the coefficient at $\frac{1}{z}$ in the
expression \ref{hol-dif-k-formula} to be zero, which ensures that the expression
\ref{hol-dif-k-formula}
 will be
holomorphic at infinity.
It is important and we will show it latter that the condition
\ref{hol-dif-k-cond} coincides with the moment map equals zero condition
for the hamiltonian reduction.
The sections of $End(\ML)\otimes \K$ by definition are expressions of the type
$\sum_i \Phi_i(z)\otimes \omega_i$, where $\Phi_i\in End(\ML); \omega_i \in \K$;
 we claim that the
expression \ref{hol-dif-k-formula} can be represented in this form.
The authors  are  unable to provide something like a simple formula
for such decomposition, but it seems that it does exist.
Let us give examples illustrating the proposition, and why it is not so
trivial.

{\Ex ~ \label{end-k-ex4} }
 Consider $Spec \{ f\in \CC[z],
f(\epsilon)=f(0) \}$, where $\ep^3=0$.
The endomorphisms of the module $\ML$  are given (see example \ref{end-ex4})
by $\Theta(z)=\Theta+ [\Lambda_1,\Theta] z+
([\Lambda_2,\Theta]+[\Theta,\Lambda_1]\Lambda_1)z^2 +z^3 \Theta(z)$,
where $\Theta, \tilde \Theta(z)$ are arbitrary. The sections of the canonical module
are given by $\frac{c_1 dz}{z^2}+ \frac{c_2 dz}{z^3} +c(z)dz$,
where $c(z)$ is holomorphic.
Hence the sections of $End(\ML)\otimes \K$ can be described as
\begin{equation*}
\begin{split}
 \Phi(z)&=(\Theta_{1}+ [\Lambda_1,\Theta_1] z+
([\Lambda_2,\Theta_1]+[\Theta_1,\Lambda_1]\Lambda_1)z^2) \frac{dz}{z^2}\\ &+
(\Theta_{2}+ [\Lambda_1,\Theta_2] z+
([\Lambda_2,\Theta_2]+[\Theta_2,\Lambda_1]\Lambda_1)z^2) \frac{dz}{z^3}
+\mbox{{\it holomorphic in z terms}}\\ &
= \frac{\Theta_2 dz}{z^3}+\frac{(\Theta_1+[\Lambda_1, \Theta_2])dz}{z^2}+
\frac{([\Theta_2,\Lambda_1]\Lambda_1
+[\Lambda_2, \Theta_2]+[\Lambda_1, \Theta_1])dz}{z} 
\\ &+\mbox{{\it holomorphic in z terms}},
\end{split}
\end{equation*}
for some matrices $\Theta_1, \Theta_2$.
On the other hand let us consider the receipt given in proposition above:
\begin{equation*}
\begin{split}
\Phi (z)&=
(Res_{\ep} \frac{\Phi(\ep)dz}{z-A(\ep)} -
\frac{\Lambda^{-1}(\ep)\Phi(\ep)\Lambda(\ep)dz}{z-B(\ep)})\\ &=
\frac{\Phi_2 dz}{z^3}+\frac{\Phi_1 dz}{z^2}+\frac{
[\Lambda_1,\Phi_1]+[\Lambda_2,\Phi_2]+\Lambda_1[\Phi_2,\Lambda_1] )
dz}{z}.
\end{split}
\end{equation*}
To prove our proposition in this example we should rewrite the expression
above in terms of $\Theta_i.$ We see that there are  three equations for the only two
parameters $\Phi_1,\Phi_2$, but one can show that nevertheless the third equation
is dependent of the first two.
It means that if we  take $\Theta_2=\Phi_2$,
$\Theta_1=\Phi_1-[\Lambda_1,\Phi_2]$ then the coefficients at
$\frac{dz}{z^3},\frac{dz}{z^3}$ are as they should be and
the coefficient at $\frac{dz}{z^3}$ is automatically the
necessary one.
So let us say again what we have get:
the expression $$\Phi (z)=
(Res_{\ep} \frac{\Phi(\ep)dz}{z-A(\ep)} -
\frac{\Lambda^{-1}(\ep)\Phi(\ep)\Lambda(\ep)dz}{z-B(\ep)})$$
can be represented in the form
$$ \Theta_{1}(z) \frac{dz}{z^2}+
\Theta_2(z) \frac{dz}{z^3}
+\mbox{{\it holomorphic in z terms}}, $$ where
$$\Theta_i(z)=(\Theta_{i}+ [\Lambda_1,\Theta_i] z+
([\Lambda_2,\Theta_i]+[\Theta_i,\Lambda_1]\Lambda_1)z^2)$$ are
elements of $End(\ML)$. To do this one puts $\Theta_2=\Phi_2$,
$\Theta_1=\Phi_1-[\Lambda_1,\Phi_2]$. Thus $\Phi(z)$ is a section
of $End(\ML)\otimes \K$.

{\Ex ~ \label{end-k-ex3} }
 Consider the cusp curve $Spec \{ f\in \CC[z],
f(\epsilon)=f(0) \}$, where $\ep^2=0$.
The endomorphisms of the module $\ML$  are given (see example \ref{end-ex3})
by $\Theta(z)=\Theta+ [\Lambda_1,\Theta] z+z^2 \tilde \Theta(z)$,
where $\Theta, \tilde \Theta(z)$ are arbitrary. The sections of the canonical module
are given by $\frac{c_1 dz}{z^2}+c(z)dz$,
where $c(z)$ is holomorphic.
So the sections of $End(\ML)\otimes \K$ can be described as
$$ \Phi(z)=(\Theta + [\Lambda,\Theta z]) \frac{dz}{z^2}
+\mbox{{\it holomorphic in z terms}}.$$
The global sections $H^0(End(\ML)\otimes \K)$
are such $ \Phi(z)=(\Theta + [\Lambda,\Theta ](z) ) \frac{dz}{(z)^2}$,
which are regular at infinity, hence $[\Theta, \Lambda]=0$ and the global sections are:
$ \Phi(z)=(B-A) \frac{ \Theta  dz }{(z)^2} $, where $\L \Theta=\Theta \L$.
On the other hand let us consider $\Phi (\ep)=\Phi_1\frac{1}{\ep^2} $
and consider
the receipt given in proposition above:
$$
\Phi (z)=
(Res_{\ep} \frac{\Phi(\ep)dz}{z-A(\ep)} -
\frac{\Lambda^{-1}(\ep)\Phi(\ep)\Lambda(\ep)dz}{z-B(\ep)})=
\frac{\Phi_1 dz}{z^2}+\frac{ [\Lambda_1,\Phi_1] dz}{z}.$$
We see that in this case the two expressions coincide.

{\Ex ~ \label{end-k-ex21} }
Consider the node curve $Spec \{ f\in \CC[z],
f(A)=f(B) \}$.
The endomorphisms of the module $\ML$  are given (see example \ref{end-ex21})
by $$\Theta(z)=\Theta (z-A) - \Lambda\Theta\Lambda^{-1} (z-B) + (z-A)(z-B)\tilde \Theta(z),$$
where $\Theta, \tilde \Theta(z)$ are arbitrary. The sections of the dualizing module
are given by $$\frac{c dz}{z-A}+\frac{cdz}{z-B}+holomorphic~in~z.$$
Hence the sections of $End(\ML)\otimes \K$ can be described as
$$ \Phi(z)=(B-A)( \frac{ \Lambda \Theta \Lambda^{-1} dz }{z-A} -
\frac{ \Theta dz }{z-B})+holomorphic~in~z.$$
So the global sections $H^0(End(\ML)\otimes \K)$
are such $$ \Phi(z)=(B-A)( \frac{ \Lambda \Theta \Lambda^{-1} dz }{z-A} -
\frac{ \Theta dz }{z-B}),$$ which are regular at infinity, hence $\L \Theta \Lambda^{-1}  -
\Theta=0$ and the global sections are:
$$ \Phi(z)=(B-A)( \frac{ \Theta  dz }{z-A} -
\frac{ \Theta dz }{z-B}),$$ where $\L \Theta=\Theta \L$.
On the other hand let us consider $\Phi (\ep)=\frac{\Phi}{\ep} $
and consider
the receipt given in proposition above:
$$
\Phi (z)=
(Res_{\ep} \frac{\Phi(\ep)dz}{z-A} -
\frac{\Lambda^{-1}\Phi(\ep)\Lambda dz}{z-B})=
\frac{\Phi dz}{z-A}+\frac{ \Lambda^{-1}\Phi \Lambda dz}{z-B}.$$
So we see that in this case the two expressions coincide up to
$\Theta=(B-A)\Lambda \Phi
\Lambda^{-1}.$

\subsection{$H^1(End(\ML))$}

By definition $\ML$ is a submodule of $\CC[z]^{r}$.
So we have an exact sequence of modules
$\ML\mapsto \CC[z]^{r}\mapsto \CC^{p}$ over an algebra
$\{ f\in \CC[z], f(A(\epsilon))=f(B(\epsilon)) \}$,
where $\ep^N=0.$ $\CC[z]^{r}\mapsto \CC^{p}$
is a resolution for $\ML$. The complex of endomorphisms of a resolution
gives a resolution of endomorphisms of $\ML$. This is one way to argue
the proposition below, another is to use \v{C}ech description of
$H^1(End(\ML))$. We will prove the proposition by the second method (see below).

{\Prop
\label{lem-chi}
The space of matrix polynomials
$\chi(z)=\sum_{i=0}^{N-1} \chi_i z^i$ considered as endomorphisms of $\CC[z]^{r}$,
maps surjectively to $H^{1}(End(\ML))$.
The kernel of this map consists of the sum of the two linear subspaces
in the space of matrix polynomials $\chi(z)$:
the first space is the space of constant polynomials $\chi(z)=\chi_0$
and the second space consists of such matrix polynomials
which satisfy the condition:
$\chi(A(\ep))=\Lambda(\ep)\chi(B(\ep))\Lambda(\ep)^{-1}$.
(Let us mention that the intersection of these two subspaces is
precisely $H^{0}(End (\ML))$ and the second subspace consists of such $\chi(z)$ which
gives endomorphisms of the module $\ML$ over the affine chart without infinity.)
}
\\
{\bf Proof~ } The proposition is quite obvious from the
point of view of \v{C}ech description of
$H^1(End(\ML))$. Let us cover our singular curve
by two charts $U_{P} = \Sigma \backslash \infty$,
$U_{\infty}=\Sigma \backslash  P$, where $P$ is the singular point.
Then $$H^1(End(\ML))= End(\ML) (  U_{P} \bigcap U_{\infty})
/  End(\ML) (  U_{P})  \oplus End(\ML) (U_{\infty}).$$
As we know from proposition \ref{end-prop}
$End(\ML) (  U_{P}) $ consists in polynomials
$\chi(z)$, which satisfy
$\chi(A(\ep))=\Lambda(\ep)\chi(B(\ep))\Lambda(\ep)^{-1}.$ And this proves
the proposition.

{\Rem ~ } For the node curve $f(A)=f(B)~ (A,B \in \CC)$ the summation in the formula
$\chi(z)=\sum_{i=0}^{N-1} \chi_i z^i$ must be understood in the interval
$i=0,1$ (i.e.  one should consider
$\chi(z)=\chi_0+\chi_1 z$), the same convention about summation
in the case of the node curve should be accepted everywhere below
(see example \ref{ex-tang-lamb-1} for details.)


{\Ex ~ \label{ex-tang-lamb-0}} in the abelian case (i.e. when $\ML$ is a rank
$1$ module) for any $\Lambda $ it is known that $End(\ML)$ is just $\mathcal{O}$.
So in the abelian case the proposition claims that
$H^1(\mathcal{O})$ is the factor space of the space of all polynomials
$\sum_{i=0}^{N-1} \chi_i z^i$ by the space of polynomials
$\chi(z)$ which satisfy the condition $\chi(A(\ep))=\chi(B(\ep))$.
This is obviously true. For example this can be seen from the
exact sequence: ${\mathcal{O}}\to {\mathcal{O}}^{norm}\to {\CC_{P}} $,
which gives that: $H^1(\mathcal{O})=H^0(\ {\CC_{P}} )$.


%
%
%

It's well-known that the vector space
$H^{1}(End(\ML))$ is the tangent space to deformations of $\ML$ as an algebraic
vector bundle, on the other hand we know that all vector bundles are given by
$\Lambda(\ep)$, so taking some $\delta\chi(z)$, where $\delta^2=0$,
$\chi(z)=\chi_i z^i$ we must find the corresponding deformation of
$\ML.$ $\delta\chi(z)$ should deform
$\ML$ to some $M_{\tilde\Lambda(\ep)}$, where $\tilde\Lambda(\ep)=
\Lambda(\ep) +\delta \Delta_{\Lambda(\ep)}$. Our aim is to determine
$\Delta_{\Lambda(\ep)}$:

{\Prop \label{chi-deform-lamb}
Matrix polynomial $\chi(z)=\sum_{i=0,...,N-1} \chi_i  z^i$, which is considered as an element
of $H^{1}(End(\ML))$ due to the proposition above, gives the following
deformation of $\Lambda(\ep)$:
\bea
\delta_{\chi(z)} \Lambda(\ep)=
\chi(A(\ep))\Lambda(\ep)- \Lambda(\ep)\chi(B(\ep)).
\eea
}

{\Rem ~ }  One knows that $H^{2}(Coherent ~  sheaves)=0$ for the case
of curves and so  by   general theory
 the map from $H^1(End(\ML))$ to the tangent space of deformations of the
bundle $\ML$ is a bijection. The formula above can be taken as a definition
for the map from the space of matrix polynomials $\chi(z)=\sum_i \chi_i z^i$
to the space $H^1(End(\ML)).$ It means (it could be taken as a definition) that
we associate with the matrix polynomial $\chi(z)=\sum_i \chi_i z^i$
such an element of $H^1(End(\ML))$ which deforms the bundle $\ML$
by the formula
$\Lambda \mapsto \Lambda(\ep) +
\chi(A(\ep))\Lambda(\ep)- \Lambda(\ep)\chi(B(\ep))$.
What must be  proved after such a definition is that one must describe
the Serre's pairing between $H^0(End (\ML) \otimes \K)$ and
the $H^1(End (\ML))$.
We describe Serre's pairing in  proposition
\ref{lem-serre-pair}. The essential point consists in the
interplay: if $\chi(z) $  acts on $\Lambda$  as above then
the Serre's pairing is described as in proposition \ref{lem-serre-pair}.

{\Cor
The matrix polynomial $\chi(z)=\chi_i z^i$,
which satisfy the condition
$\chi(A(\ep))=\Lambda(\ep)\chi(B(\ep))\Lambda(\ep)^{-1}$
does not change $\Lambda$. This fact is in full
agreement with the proposition \ref{lem-chi} which
says that such polynomial gives zero element
in $H^1(End(\ML))$.
}
{\Cor
The matrix polynomial $\chi(z)=\chi_0$,
conjugates $\Lambda$ by the constant matrix, so
it gives the same vector bundle. This fact is in
full agreement with the proposition \ref{lem-chi} which says that such polynomial
gives zero element in $H^1(End(\ML))$.  }

The proposition may be argued as follows:
consider an element $ 1+\delta\chi(z)$, where $\delta^2=0$,
it is an infinitesimal automorphism of the module $\CC[z]^{r}$ corresponding
to the endomorphism $\chi(z)$.
Having such an automorphism it is clear how to deform the
module $\ML$:  new module is the set of elements
$(1+\delta\chi(z)) s(z)$, where $s(z)$ is an element of $\ML$.
The elements of the type $\tilde s(z)=(1+\delta\chi(z)) s(z)$
obviously satisfy the condition:
$$\tilde s(A(\ep))= (1+\delta\chi(A(\ep)))
\Lambda(\ep)(1+\delta\chi(B(\ep)))^{-1}
\tilde s(B(\ep)),$$ hence\\
$$\tilde s(A(\ep))= (\Lambda(\ep)+\delta(\chi(A(\ep))\Lambda(\ep)-
\Lambda(\ep)\delta\chi(B(\ep)))) \tilde s(B(\ep))$$ and
we see that the new module
is the module $M_{\Lambda(\ep)+\delta(\chi(A(\ep))\Lambda(\ep)-
\Lambda(\ep)\delta\chi(B(\ep)))}$. Q.E.D.

There is another formal argument: the module $\ML$
is embedded in the module $\CC[z]^r$, this module cannot be deformed,
so deformations of $\ML$ are governed only by the deformation of the
embedding $\ML \to \CC[z]^r$. The elements of $H^1(End (\ML))$ are
identified with some elements of $End(\CC[z]^r)$
because of the fact that  $\CC[z]^r\to \CC^p$ is a resolution
of $\ML$. The elements of $End(\CC[z]^r)$ obviously acts on the set of embeddings
$\ML\to \CC[z]^r $.

One can argue the proposition above more formally
using the \v{C}ech description
of cohomologies and sheaves. We consider the covering
of our projective curve consisting of two charts:
first is everything except the singular point, the second
chart is not really an honest open set but a limit
of the open sets - infinitesimal neighborhood of singular point
i.e. $Spec \{ f\in
\CC(z), f$ is regular at $a_0$ and $b_0$ and $f(A(\ep))=f(B(\ep)) \}$.
(We need to consider such infinitesimal neighborhood
because only in such neighborhood of singular point all modules
become trivial, because it is a spectra of a local ring.)
Hence any module on a singular curve can be given by glueing
the two trivial modules by the gluing function on the intersection
of the two charts. Intersection of these two charts is the "general point"
i.e. $Spec \{ \CC(z)\}$.
The first task is to describe the module $\ML$ by the
gluing function.
After that it's obvious how to calculate what
deformation corresponds to elements of $H^{1}(End(\ML))$.
We represent an element of $H^{1}(End(\ML))$ as an element
of $\chi\in End(\ML)$ on the intersection of two charts and
one should simply multiply the gluing function by the
element $1+\delta \chi$. We obtain the new glueing function and the
new bundle. It can be again represented in the form $\ML$.
This gives the same results as above.

Let us give examples  illustrating
proposition \ref{lem-chi} and proposition \ref{chi-deform-lamb}.

{\Ex ~ \label{ex-tang-lamb-1}}
Consider the node curve with the affine part
$Spec \{ f(A)= f(B) \}$, where $A,B \in \CC$.
Consider the matrix polynomial $\chi(z)=\chi_0+\chi_1 z + (z-A)(z-B)\tilde \chi(z).$
According to proposition \ref{chi-deform-lamb}
it acts on $\Lambda$ by the formula
$\delta_{\chi}\Lambda = \chi(A)\Lambda-\Lambda \chi(B).$
The part $(z-A)(z-B)\tilde \chi(z)$
is zero in $A,B$ and it does not act on $\Lambda$.
We can only consider $\chi(z)=\chi_0+\chi_1 z$.
According to proposition \ref{lem-chi}
$H^1(End(\ML))$ is the factor of the space $\chi_0+\chi_1 z$
by the sum of the spaces $\chi(z)=\chi_0$ and $\chi(z)=\Theta(z-A)-
\Lambda \Theta \Lambda ^{-1}(z-B) $,where $\chi_0$, $\Theta$ are
arbitrary matrices. It would be nice
to have explicit parameterization of the orthogonal
(with respect to Killing form) compliment
to the sum of these two  subspaces  and to generalize it to the case of
schematic points. The intersection of these two subspaces
is the subspace of constant matrices $\chi(z)=\chi_0$, such that $\chi_0$
commutes with $\Lambda$ - this intersection is  $H^0(End(\ML))$. We see that
$dim H^1 (End(\ML))= dim H^0(End(\ML))$.
This precisely coincides with the  calculation from the  Riemann-Roch
theorem: $$dim H^0 (End(\ML))- dim H^1(End(\ML))= deg (End (\ML))
- n^2 (1-dim H^1( \mathcal{O)})=0.$$

{\Ex ~ \label{ex-tang-lamb-cusp} }
Consider the cusp  curve $\Sigma^{proj}$.
Recall that the affine part of $\Sigma^{proj}$ is given by
$ Spec \{ f(z) \in \CC[z]: f(P+\ep)=f(P) \}$,
the bundle $\ML$ corresponds to the module which is
$\{ s(z)\in \CC[z]^r: s(P+\ep)=(1+\L\ep) s(P)\}$ over the affine
part.
For the cusp curve proposition \ref{lem-chi} explicitly means that:
the space of matrix polynomials
$\chi(z)= \chi_0+\chi_1 (z-P)$ maps surjectively to $H^{1}(End(\ML))$.
The kernel of this map consists of the sum of the two linear subspaces
in the space of matrix polynomials $\chi(z)$:
the space of  constant polynomials $\chi(z)=\chi_0$
and the space of such matrix polynomials
which satisfy the condition:
$\chi_1=[\Lambda, \chi_0]$, for $i=2,...,N$.
For the cusp curve proposition \ref{chi-deform-lamb} explicitly means
that:
matrix polynomial $\chi(z)= \chi_0+\chi_1 (z-P)$ gives the following
deformation of $\Lambda$:
 $$\delta_{\chi(z)} \Lambda= \Lambda + \chi_1+[\chi_0,\L].
$$

Let us give an example (which seems a little amusing for us) illustrating
lemma \ref{lem-chi} and proposition \ref{chi-deform-lamb}.

{\Ex ~ \label{tang-lamb} }
Consider the curve with the affine part
$Spec \{ f(\ep)= f(B(\ep)) \}$, where
$\underbrace{B(B(B...(B}_{\mbox{k times}} (\ep)))=\ep$,
 $\ep^N=0$.
According to theorem \ref{bund-on-sigm} the vector
bundles are given by such $\Lambda(\ep)$ that
\bea \label{lamb-condition}
\Lambda(\ep)\Lambda(B(\ep))\Lambda(B(B(\ep)))...\Lambda
(\underbrace{B(B(B...(B}_{\mbox{k-1 times}} (\ep))))=Id
\mbox{ and } \Lambda_0=Id.
\eea
Consider an arbitrary matrix valued polynomial $\chi(z)=\sum_{i=0,...N-1}
\chi_i z^i$.
Consider $$\tilde \Lambda (\ep) =
\Lambda(\ep)+ \delta (\chi(\ep)\Lambda(\ep)-\Lambda(\ep)\chi(B(\ep))),$$ where
$\delta^2=0$.
Then $\tilde \Lambda $ again satisfies the condition \ref{lamb-condition}.

This can be seen simply rewriting
$$\tilde \Lambda(\ep) = exp(\delta \chi(\ep)) \Lambda(\ep)
exp(-\delta \chi(B(\ep))).$$

We can restate the example above as saying that the tangent
space to the space of $\Lambda(\ep)$ at the point
$\Lambda_0(\ep)$ which is defined by the
equation \ref{lamb-condition} is the quotient of the space of
all $\chi(z)$ by the space $\chi(\ep)=\Lambda_0(\ep)
\chi(B(\ep))\Lambda_0(\ep)^{-1}$.

%

\vskip 1cm

Let us describe the  Serre's duality.

{\Prop \label{lem-serre-pair}
The Serre's pairing between $H^{0}(End(\ML)\otimes \K)$ and
$H^{1}(End(\ML))$
can be written in terms of matrix polynomials  $\tilde \Phi(z)$
and  $\chi(z)$ very simply:
\bea
{\label{pairing-1-0}}
Tr Rez_{z} \chi(z) \tilde \Phi(z).
\eea
}

{\Cor Consider a matrix polynomial $\chi(z)=\sum_i \chi_i z^i$,
which satisfy the condition
$\chi(A(\ep))=\Lambda(\ep)\chi(B(\ep))\Lambda(\ep)^{-1}$.
The Serre's pairing (given by the formula
\ref{pairing-1-0})  between such $\chi(z)$ and
arbitrary $\Phi(z)\in H^{0}(End(\ML)\otimes K)$
is identically zero.
This fact is in a full
agreement with proposition \ref{lem-chi} which
says that such polynomial gives the zero element
in $H^1(End(\ML))$.
}

{\Cor
Consider a matrix polynomial $\chi(z)=\chi_0$,
The Serre's pairing (given by the formula
\ref{pairing-1-0}) between such $\chi(z)$ and
arbitrary $\Phi(z)\in H^{0}(End(\ML)\otimes \K)$
 is identically zero.
This fact is in a
full agreement with proposition \ref{lem-chi} which says that such polynomial
gives the zero element in $H^1(End(\ML))$.  }


The lemma is quite obvious.
To prove corollaries we use proposition
\ref{hol-dif-k} in order to  represent $\tilde \Phi(z)$ as
a matrix polynomial:
$$Res_{\ep} \frac{\Phi(\ep)dz}{z-A(\ep)} -
\frac{\Lambda^{-1}(\ep)\Phi(\ep)\Lambda(\ep)dz}{z-B(\ep)}$$
for some $\Phi(\ep)$. And the corollaries follow obviously
from the properties that $Tr AB=Tr BA$ and $Res_{z}Res_{\ep}=
Res_{\ep}Res_{z}$.

{\Rem ~ }
To prove the second corollary we should also use the condition
$$Res_{\ep} (\Phi(\ep)-\Lambda^{-1}(\ep)\Phi(\ep)
\Lambda(\ep))=0.$$
The first corollary does not use this condition for $\Phi(\ep)$
and is true for all $\Phi(z)$ represented in the form
$$Res_{\ep} \frac{\Phi(\ep)dz}{z-A(\ep)} -
\frac{\Lambda^{-1}(\ep)\Phi(\ep)\Lambda(\ep)dz}{z-B(\ep)}$$
with arbitrary $\Phi(\ep)$.


\section{Symplectic form}
\subsection{Description
 \label{sect-1-form}
}

In the expert's language in this section we prove the following:

{\bf Claim: } {\it
the canonical one-form on the cotangent bundle to the  moduli space of
vector bundles on the curve $\Sigma_{A(\ep), B(\ep)}^{proj}$
in terms of $\Lambda(\ep), \Phi(\ep)$
coincides with the form:
\bea
Tr Res_{\ep} \Lambda(\ep)^{-1} \Phi(\ep)d \Lambda(\ep).
\eea
}

Let us formulate the claim above more exactly.
In proposition \ref{hol-dif-k} we showed that the  elements of
$ H^{0}(End_{\ML}\otimes \K)$ can be described as
$$\tilde \Phi(z)=
 \frac{\Phi(\ep)dz}{z-A(\ep)} -
\frac{\Lambda^{-1}(\ep)\Phi(\ep)\Lambda(\ep)dz}{z-B(\ep)}
$$ where $\Phi(\ep)$ is such that it satisfies the
condition:
\bea Res_{\ep} \Phi(\ep)
-\Lambda^{-1}(\ep)\Phi(\ep)\Lambda(\ep)=0.
\label{holom-cond}
\eea
The space $H^{0}(End_{\ML}\otimes \K)$ is cotangent to the moduli space
of vector bundles at the point $\ML$.
By proposition \ref{hol-dif-k} we have the map $p$
\bea
p: \{ \mbox{set of pairs:} (\Lambda(\ep),\Phi(\ep))
\mbox{which satisfy the condition \ref{holom-cond}} \}
\to T^*Bun.
\label{map-p}
\eea
More exactly we also request $\Lambda(\ep)$ to satisfy the
condition as in theorem \ref{bund-on-sigm}.
The map $p$ is not one-to-one but it is surjective.
The pairs conjugated by the action of $GL(n)$
gives the same point in $T^*Bun$.
We will show latter that the condition
\ref{holom-cond} is the same as to say that "the moment map equals zero"
for the natural action of $GL(n)$ and the canonical symplectic form
on the space
$(\Lambda(\ep),\Phi(\ep))$. (This space
is cotangent to the space $(\Lambda(\ep))$).

Consider some pair $\Lambda(\ep),\Phi(\ep)$
which satisfies the condition \ref{holom-cond}.
Consider some tangent vector $\delta {\Lambda}(\ep), \delta {\Phi}(\ep)$,
such that it is tangent to the surface defined by the equation
\ref{holom-cond}.
{\Th
\label{th-1-form}
The canonical one-form on the cotangent bundle to the
moduli space of vector bundles applied to the vector
$p(\delta {\Lambda}(\ep), \delta {\Phi}(\ep))$ at a point
$p(\Lambda(\ep),\Phi(\ep))$ is equal to
\bea
Tr Res_{\ep} \Lambda(\ep)^{-1} \Phi(\ep) \delta {\Lambda}(\ep).
\eea
}
\\
{\bf Proof~}
Let us take the tangent vector $\delta_{\chi(z)}$ considered
in propositions \ref{lem-chi},\ref{chi-deform-lamb}.
Consider the tangent vector $p(\delta_{\chi(z)}, \delta {\Phi})$
with arbitrary $\delta {\Phi}$ at arbitrary point
$p(\Lambda(\ep),\Phi(\ep))$.
By definition of the canonical $1$-form on the
cotangent bundle to the manifold its value on the vector
$p(\delta_{\chi(z)}, \delta {\Phi})$ at the point
$p(\Lambda(\ep),\Phi(\ep))$ equals to
\bea
< p(\Phi(\ep)) | p(\delta_{\chi(z)}) > \nn\\
\mbox{by propositions \ref{hol-dif-k},\ref{lem-serre-pair} equals to:} \nn\\
Tr Res_z
(Res_{\ep} \frac{\Phi(\ep)dz}{z-A(\ep)} -
\frac{\Lambda^{-1}(\ep)\Phi(\ep)\Lambda(\ep)dz}{z-B(\ep)}) \chi(z)=\nn\\
=
Tr
Res_{\ep} {\Phi(\ep)}{\chi(A(\ep))} -
{\Lambda^{-1}(\ep)\Phi(\ep)\Lambda(\ep)}{\chi(B(\ep))} =\nn\\
=
Tr
Res_{\ep} \Lambda^{-1}(\ep)\Phi(\ep) \Bigl({\chi(A(\ep))} \Lambda(\ep) -
{\Lambda(\ep)}{\chi(B(\ep))} \Bigr)  =\nn\\
\mbox{by proposition \ref{chi-deform-lamb} it equals to the desired result:
} \nn\\
=Tr Res_{\ep} \Lambda(\ep)^{-1} \Phi(\ep) \delta_{\chi(z)}.
\eea
By proposition \ref{lem-chi} all the tangent vectors can
be represented as some $\delta_{\chi(z)}$ and the theorem is proved
$\blacksquare$

\subsection{Hamiltonian reduction, moment map and holomorphity condition
}

{\Lem The action of the group $GL(n)$ by conjugation on pairs
$\L(\ep),\Phi(\ep)$ is hamiltonian with respect to the symplectic
form $Tr Res_\ep d (
\Lambda(\ep)^{-1} \Phi(\ep)) \wedge d \Lambda(\ep)$
and the moment map is given by the formula:
\bea \label{mom-map}
Res_{\ep} \left(\Phi(\ep)-\Lambda(\ep)^{-1}\Phi(\ep)\Lambda(\ep)\right)
\eea
}

The proof is standard and is left to the reader.

{\Cor we see that the equation "moment equals zero" on $\Phi(\ep)$
coincides with the holomorphity condition for
$\Phi(z)\in H^0(End(\ML\otimes K))$ see equation (\ref{holom-cond}).
}

As a corollary we obtain the following theorem:

{\Th
\label{red-th}
The phase space of Hitchin system
which is the total space of the cotangent bundle to the moduli space
of vector bundles on a singular curve can be obtained as a hamiltonian
reduction of the space of pairs $\L(\ep),\Phi(\ep)$ with the symplectic form
$Tr Res_\ep d (
\Lambda(\ep)^{-1} \Phi(\ep)) \wedge d \Lambda(\ep)$
 by the  action the of $GL(n)$ by conjugation.
(More precisely one should speak about submanifold in the
space $\L(\ep),\Phi(\ep)$ as in theorem
\ref{bund-on-sigm}).
The projection acts as follows: $\L(\ep)$ maps to vector the bundle $\ML$ as described in
theorem \ref{bund-on-sigm}, $\Phi(\ep)$ maps to
the cotangent covector as it is described in proposition
\ref{hol-dif-k} (recall that $H^0(End(\ML\otimes K))=T^*_{\ML}Bun$).
Let us emphasize that this map  respects the symplectic structures
on both spaces which means that the canonical symplectic structure on
the cotangent bundle to the moduli space of vector bundles
is precisely the one obtained from
$$Tr Res_\ep d (
\Lambda(\ep)^{-1} \Phi(\ep)) \wedge d \Lambda(\ep)$$ on the space
of pairs $\L(\ep),\Phi(\ep)$.

The Lax operator is defined as
$$\Phi(z)=Res_{\ep} \left(\frac{\Phi(\ep)dz}{z-A(\ep)} -
\frac{\Lambda^{-1}(\ep)\Phi(\ep)\Lambda(\ep)dz}{z-B(\ep)}\right).$$
The generating functions for the Hitchin's hamiltonians
are defined as $Tr\Phi(z)^k$.
}

\section{ \label{commut-sect} Integrability}
The aim of this section is to prove the commutativity of Hitchin's
hamiltonians for the case of our singular curves.
(Hitchin's proof is not applicable in this case).
In the previous section we described the Hitchin system
as the Hamiltonian reduction.
The aim of this section is to prove that
$Tr \Phi(z)^k$ and $Tr \Phi(w)^l$ Poisson commute for all $z,w,k,l$.
This functions are invariant with respect to the $GL(n)$ action,
so they can be pushed down to the reduced space, which is by definition
the phase space of Hitchin system and this functions are by definition
Hitchin's hamiltonians.
The main property of hamiltonian reduction that it preserves the
Poisson bracket between invariant functions.
So it's enough to prove that
$Tr \Phi(z)^k$ and $Tr \Phi(w)^l$ Poisson commute on the nonreduced
space, which is just the space of pairs $\L(\ep),\Phi(\ep)$.
This can be done by the $r$-matrix technique.
It's very easy and well-known among experts that the
Poisson bracket between $\Phi(\ep)$ is of $r$-matrix form,
but it is quite amusing for us that it's also true
for the $$\Phi (z)=
Res_{\ep} \left(\frac{\Phi(\ep)dz}{z-A(\ep)} -
\frac{\Lambda^{-1}(\ep)\Phi(\ep)\Lambda(\ep)dz}{z-B(\ep)}\right)$$ (see lemma
\ref{r-for-phiz}).
It would be very interesting to understand the relation between
our $r$-matrix approach in the case of singular curves and
$r$-matrix approach proposed in \cite{ER1}, \cite{VALD}.

\subsection{Truncated $\delta$-functions and their properties}

The formula for the Poisson bracket between canonically conjugated
variables $P(\eta)$ and $\Lambda(\ep)$ (see the next section) includes
expression
$$\delta_{\ep}^{\eta}=\sum_{k=0,...,N-1}\frac{(\eta)^{k}}{(\ep)^{k+1}}.$$
Informally speaking one should think about it
as about the delta-function. We will prove some elementary properties
of it which we need in order to simplify expressions for Poisson
brackets between $\Phi(\eta)$.

{\Lem For arbitrary function $f(\ep)$ and the
"delta"-function given by \bea
\delta_{\ep}^{\eta}=\sum_{k=M,...,N-1}\frac{(\eta)^{k}}{(\ep)^{k+1}}
\eea the following equality is true: \bea \label{f-delta} \left|
f(\ep) \delta_{\eta}^{\ep}\right|_{-}= \left| f(\eta)
\delta_{\ep}^{\eta}\right|_{-} \eea where $|...|_{-}$ is defined
as a linear operation, which acts as follows on the basis of monomials:
\bea \label{cut2v} \left|\frac{1}{\eta^k\ep^l}\right|_{-}=
\left\{\begin{array}{ll} \frac{1}{\eta^k\ep^l}, & \mbox{ if } N
\ge k\ge M+1 \mbox{ and } N \ge l\ge M+1 \\ 0, & \mbox{ otherwise
}
\end{array} \right.
\eea }{\bf Proof~} It's easy to see that for $f(\ep)=\ep^{-l}$,
$l\in \mathbb{Z}$ both sides of the equality can be rewritten as:
$$\sum_{ \max(M; l-N\le k \le \min(N-1; l-(M+1))}
\frac{1}{\eta^{k+1}\ep^{l-k}}.$$ Analogously we define
$|...|_{-}$ for the function of one variable as a linear operation
defined on the monomials as follows \bea
\left|\frac{1}{\eta^k}\right|_{-}= \left\{\begin{array}{ll}
\frac{1}{\eta^k}, & \mbox{ if } N \ge k\ge M+1 \\ 0, & \mbox{
otherwise }
\end{array} \right.
\eea Obviously $| f(\ep)g(\eta)|_{-}=| f(\ep)|_{-}| g(\eta)|_{-}$
where $|...|_{-}$ for the function of two variables is defined by
(\ref{cut2v}). We introduce another linear operation
$\left|...\right|_{\eta}$ for functions in $\eta$ as follows: \bea
\left|\eta^k\right|_{\eta}= \left\{\begin{array}{ll} \eta^k, &
\mbox{ if } -N \leq k < N \\ 0, & \mbox{ otherwise }
\end{array} \right.
\eea We will use this cutting only for polar expressions and in
this case it is equivalent to the multiplication in $\C[\frac 1
{\eta}]/{\frac 1 {\eta}}^{N+1}.$ {\Cor \label{cor1} For the case
$M=0$, $f(\ep)=\sum_{-N\leq i<N-1} f_i \ep^i$, $g(\ep)=\sum_{0\leq
i < N} g_i \ep^i$ the following is true: \bea\left| f(\ep) g(\eta)
\delta_{\eta}^{\ep}\right|_{-}= \left|\left|
f(\ep)\delta_{\eta}^{\ep}\right|_{-}g(\eta)\right|_{-}=
 \left|\left|
f(\eta)\delta_{\ep}^{\eta}\right|_{-}g(\eta)\right|_{-}=
 \left| f(\eta) g(\eta)
\delta_{\ep}^{\eta}\right|_{-}\eea }



{\Rem ~} In the case $M=-N$ obviously
$\delta_{\ep}^{\eta}=\delta_{\eta}^{\ep}$. In the case $M=-\infty,
N=+\infty$ we do need any cutting in the formula \ref{f-delta}.
And we obtain the well-known formula: $f(\ep) \delta= f(\eta)
\delta$.

{\Ex ~ } Let $f(\ep)=\ep^{-4}$, and
$\delta_{\ep}^{\eta}=\sum_{k=0,...,N-1}\frac{(\eta)^{k}}{(\ep)^{k+1}}$,
where $N\geq 1$, then both sides of the equality \ref{f-delta}
gives: $\frac{1}{\ep^3\eta^2} +\frac{1}{\ep^2\eta^3}$.

{\Lem For $f(z)=\sum_{i\in \mathbb{Z}} f_i z^i$ one obviously has:
\bea Res_{z} f(z) \delta^{w}_{z}=\sum_{i=M,...,N-1} f_i w^i \eea }

{\Lem \label{pos} For $M=0$ and $f(\ep)=\sum_{-N \leq i<0}f_i
\ep^i$ one has: $$\left|(f(\ep))-f(\eta))\delta_{\eta}^{\ep}-
|(f(\ep))-f(\eta))\delta_{\eta}^{\ep}|_{-}\right|_{\eta}=0.$$
}{\bf Proof~} The expression in question is a positive in $\ep$
part of $(f(\ep)-f(\eta))\delta_{\eta}^{\ep}.$ Let us test it for
basic monomials: $$\left|(\ep^{-n}-\eta^{-n})\delta_{\eta}^{\ep}-
|(\ep^{-n}-\eta^{-n})\delta_{\eta}^{\ep}|_{-}\right|_{\eta}=
\sum_{k=n}^{N-1}\frac {\ep^{k-n}}{\eta^{k+1}}-
\sum_{k=0}^{N-1-n}\frac {\ep^{k}}{\eta^{k+1+n}}$$ which is zero by
a simple change of indexes. $\square$

Here we present the final version of the previous lemma which is
also demonstrated by straightforward calculation. {\Lem
\label{ful}For $M=0$ and $f(\ep)=\sum_{-N \leq i< N-1}f_i \ep^i$
one has: $$\left|(f(\ep))-f(\eta))\delta_{\eta}^{\ep}-
|(f(\ep))-f(\eta))\delta_{\eta}^{\ep}|_{-}\right|_{\ep,\eta}=0.$$}

\subsection{Non-reduced Poisson bracket and $r$-matrix}

 In what follows we use
the notion of the reduced $\delta$-function with $M=0$ and $N$ is
defined by the construction of the curve, so that schematic points
are morphisms to $\C[\ep]/{\ep^N}.$ So by $\delta_\ep^\eta$ we
mean $$\delta_\ep^\eta=\sum_{k=0}^{N-1} \frac {\eta^k}
{\ep^{k+1}}.$$ Now we calculate Poisson bracket for the
ingredients of our construction. Let $\mathcal{L}$ be the space of
$\Lambda$ subject to the relation "defining the module". It is
the subset of invertible elements in $Mat_{K\times K}[\ep]/{\ep^N}.$
The cotangent bundle
$\mathcal{T}^*\mathcal{L}$ has a canonical parameterization with
conjugated variables $P(\ep)$ and $\Lambda(\ep)$ such that  the
symplectic form is defined as
\begin{equation}\label{simform}
  \omega= Res_{\ep}Tr(d P(\ep)\wedge d\Lambda(\ep)),
\end{equation}
where $$\Lambda(\ep)=\sum_{k=0}^{N-1}\Lambda_k \ep^k\qquad
P(\ep)=\sum_{k=0}^{N-1}P_k\ep^{-k-1}.$$

Matrix
coefficients $\Lambda_k^{ij}$ and $P_k^{ji}$ are
canonically conjugated. We could write the Poisson bracket in the matrix form
as follows:
\begin{equation}\label{PB1}
\{P(\ep)\otimes\Lambda(\eta)\}=\delta_\ep^\eta R,
\end{equation}
where $R$ is the trivial solution of the Yang-Baxter equation,
which is the transposition matrix in the tensor product, so it
acts as follows $$R (v_1\otimes v_2)=v_2\otimes v_1$$ and in terms
of matrix elements it could be expressed by
$$R=\sum_{i,j}e_{ij}\otimes e_{ji}.$$ It is worth to mention that
\begin{equation}\label{PB2}
\{P(\ep)\otimes P(\eta)\}=\{\Lambda(\ep)\otimes\Lambda(\eta)\}=0.
\end{equation}

Now we introduce variables $\Phi(\ep)$ by the formula
$$\Phi(\ep)=|\Lambda(\ep)P(\ep)|_{-},$$
(obviously it is true that
$P(\ep)=|\Lambda^{-1}(\ep)\Phi(\ep)|_{-}.$)

\begin{Lem}\label{l1}

\begin{equation}\label{PB3}
  \{\Phi(\ep)\otimes\Phi(\eta)\}=\left| [\Phi(\ep)\otimes
1,\delta_\eta^{\ep}R]\right|_{-}=\left|[\Phi(\ep)\otimes
1+1\otimes \Phi(\eta),\delta_{\eta}^{\ep}R]\right|_{\eta}.
\end{equation}

\end{Lem}
{\bf Proof~} $$\{\Phi(\ep)\otimes\Phi(\eta)\}=\{| \Lambda(\ep)
P(\ep)|_{-}\otimes| \Lambda(\eta)P(\eta)|_{-}\}=|\{\Lambda(\ep)
P(\ep)\otimes \Lambda(\eta)P(\eta)\}|_{-}=$$
$$|(\Lambda(\ep)\otimes 1)\delta_{\ep}^\eta R(1\otimes
P(\eta))-(1\otimes\Lambda(\eta))\delta_{\eta}^{\ep}R(P(\ep)\otimes
1 )|_{-}=$$ $$|(\Lambda(\ep)P(\ep)\otimes
1)\delta_{\eta}^{\ep}R|_{-}-|(1\otimes
\Lambda(\eta)P(\eta))\delta_{\ep}^{\eta}R|_{-}=$$ {\em / here we
used corollary \ref{cor1} and the following property of the
transposition operator \\$R (1\otimes A)=R (1\otimes
A)RR=(A\otimes 1)R $/}
$$\left|(\left|\Lambda(\ep)P(\ep)\right|_{-} \otimes
1)\delta_{\eta}^{\ep}R\right|_{-}-\left|(1\otimes
\left|\Lambda(\eta)P(\eta)\right|_{-})\delta_{\ep}^{\eta}
R\right|_{-}=$$ $$\left|(\Phi(\ep)\otimes
1)\delta_{\eta}^{\ep}R\right|_{-}-\left|(1\otimes
\Phi(\eta))\delta_{\ep}^{\eta} R\right|_{-}=$$ $$\left|
[\Phi(\ep)\otimes 1,\delta_\eta^{\ep}R]\right|_{-}=-\left|
[1\otimes\Phi(\ep),\delta_\eta^{\ep}R]\right|_{-}= $$ $$ \left|
[\Phi(\eta)\otimes 1,\delta_{\ep}^{\eta}R]\right|_{-}=-\left|
[1\otimes\Phi(\eta),\delta_{\ep}^{\eta}R]\right|_{-}.$$

For proving the second equality we use the same strategy as in the
infinite $N$ case \cite{BBT} but we have no more complex analysis
intuition in virtue of the formality of all the expressions. By
lemma \ref{pos} we have $$[\Phi(\ep)\otimes
1,\delta_\eta^{\ep}R]-\{\Phi(\ep)\otimes\Phi(\eta)\}=\left|[\Phi(\eta)\otimes
1,\delta_\eta^{\ep}R]\right|_{\eta}-|[\Phi(\eta)\otimes
1,\delta_\eta^{\ep}R]|_{-}.$$ The $|...|_{-}$ in the r.h.s. is
zero and using $$[\Phi(\eta)\otimes
1,\delta_\eta^{\ep}R]=-[1\otimes\Phi(\eta),\delta_\eta^{\ep}R]$$
one obtains the result. $\blacksquare$
\\
Now we introduce right-invariant vector fields:
\begin{equation}\label{psi}
  \Psi(\ep)=\left|P(\ep)\Lambda(\ep)\right|_{-};\qquad
  P(\ep)=\left|\Psi(\ep)\Lambda^{-1}(\ep)\right|_-.
\end{equation}
 By the same method
we obtain the following
\begin{Lem} \label{lem3} \begin{equation}\label{PB4}
  \{\Psi(\ep)\otimes\Psi(\eta)\}=-\left| [\Psi(\ep)\otimes
1,\delta_\eta^{\ep}R]\right|_{-}=-\left|[\Psi(\ep)\otimes
1+1\otimes \Psi(\eta),\delta_{\eta}^{\ep}R]\right|_{\eta}.
\end{equation}
\end{Lem}

\begin{Lem}
\begin{equation}\label{PB5}
  \left\{\Phi(\ep)\otimes\Psi(\eta)\right\}=0.
\end{equation}
\end{Lem}
{\Rem In general this is true due to the fact that right- and
left-invariant vector fields commute, but in our special case we
prefer to give an explicit demonstration.}
\\
{\bf Proof~}
$$\{\Phi(\ep)\otimes\Psi(\eta)\}=\left|\left\{\Lambda(\ep)P(\ep)\otimes
P(\eta)\Lambda(\eta)\right\}\right|_-$$
$$=\left|\Lambda(\ep)\otimes
P(\eta)\left\{P(\ep)\otimes\Lambda(\eta)\right\}+\left\{\Lambda(\ep)\otimes
P(\eta)\right\} P(\ep)\otimes \Lambda(\eta)\right|_-$$
$$=\left|\Lambda(\ep)\otimes
P(\eta)\delta_{\ep}^{\eta}R-\Lambda(\eta)\otimes
P(\ep)\delta_{\eta}^{\ep}R\right|_-$$ which is zero due to the
corollary \ref{cor1} $\square$
\\
Now we combine $\Phi(\ep)$ and $\Psi(\ep)$ into the expression
$$\Phi(z,\ep)=\left|\frac {\Phi(\ep)}{z-A(\ep)}-\frac
{\Psi(\ep)}{z-B(\ep)}\right|_{-\{\ep\}},$$ formal in $\ep$ and
analytic in $z,$ where the decomposition order is always implied
as $$\frac 1 {z-A(\ep)}=\left|\sum_{k=0}^{\infty}\frac
{A^k(\ep)}{z^{k+1}}\right|_{\ep}.$$  We need also for the analytic
half $\delta$-function $$\delta_w^z=\sum_{k=0}^\infty \frac
{z^k}{w^{k+1}}.$$
{\Lem \label{lem2}
\begin{equation}\label{PB6}
\left\{\Phi(z,\ep)\otimes\Phi(w,\eta)\right\}=\left|\left[\Phi(z,\ep)\otimes
1,R\right]\delta_{\eta}^{\ep}\delta_w^z\right|_{-\{\mbox{\tiny all
variables}\}}
\end{equation}}
{\bf Proof~} Using (\ref{PB3}),(\ref{PB4}),(\ref{PB5}) one obtains
$$\left\{\Phi(z,\ep)\otimes\Phi(w,\eta)\right\}=\left|\frac
{\{\Phi(\ep)\otimes\Phi(\eta)\}}
{(z-A(\ep))(w-A(\eta))}+\frac{\{\Psi(\ep)\otimes\Psi(\eta)\}}{(z-B(\ep))(w-B(\eta))}
\right|_{-\{\ep,\eta\}}$$ $$=\left|\frac {\left|[\Phi(\ep)\otimes
1,\delta_{\eta}^{\ep}R]\right|_{-\{\ep\}}}
{(z-A(\ep))(w-A(\eta))}-\frac{\left|[\Psi(\ep)\otimes 1
,\delta_{\eta}^{\ep}R] \right|_{-\{\ep\}}} {(z-B(\ep))(w-B(\eta))}
\right|_{-\{\ep,\eta\}}$$ /using the fact that $\frac 1
{(z-A(\ep))(w-A(\eta))}$ has only positive powers of $\ep$ we
continue/ $$=\left|\frac {[\Phi(\ep)\otimes
1,\delta_{\eta}^{\ep}R]}
{(z-A(\ep))(w-A(\eta))}-\frac{[\Psi(\ep)\otimes 1
,\delta_{\eta}^{\ep}R]} {(z-B(\ep))(w-B(\eta))}
\right|_{-\{\ep,\eta\}}$$ /and using one more time corollary
\ref{cor1} we find/
 $$=\left|\frac {[\Phi(\ep)\otimes
1,\delta_{\eta}^{\ep}R]}
{(z-A(\ep))(w-A(\ep))}-\frac{[\Psi(\ep)\otimes 1
,\delta_{\eta}^{\ep}R]} {(z-B(\ep))(w-B(\ep))}
\right|_{-\{\ep,\eta\}}$$
\\
Here we need for some properties of formal expression $\frac 1
{(z-A(\ep))(w-A(\ep))}$ $$\frac 1 {z-A(\ep)}\frac 1
{w-A(\ep)}=\left|\sum_{k=0}^{\infty}\frac
{A^k(\ep)}{z^{k+1}}\right|_{\ep}\left|\sum_{l=0}^{\infty}\frac
{A^l(\ep)}{w^{l+1}}\right|_{\ep}=\left|\sum_{m=0}^{\infty}A^{m}
(\ep)\sum_{k=0}^{m}\frac 1 {z^{k+1}w^{m-k+1}}\right|_{\ep} $$
$$=\left|\sum_{m=0}^{\infty}\frac {A^m(\ep)}{w^{m+1}}\sum_{k=0}^m
\frac {w^k}{z^{k+1}}\right|_{\ep}=\left|\left|\frac 1
{w-A(\ep)}\delta_z^w\right|_{\ep}\right|_{-\{z,w\}}=\left|\left|\frac
1 {z-A(\ep)}\delta_w^z\right|_{\ep}\right|_{-\{z,w\}}.$$ Using
this we proceed by
$$\left\{\Phi(z,\ep)\otimes\Phi(w,\eta)\right\}=\left|\left[\left(\frac
{\Phi(\ep)}{z-A(\ep)}+\frac {\Psi(\ep)}{z-B(\ep)}\right)\otimes
1,R\right]\delta_{\eta}^{\ep}\delta_w^z\right|_{-\{\ep,\eta,z,w\}}$$
and this finishes the proof because $\delta_{\eta}^{\ep}$ has only
positive powers in $\ep~ \square$
\\
The geometrical object, i.e. the cotangent vector to the space of
holomorphic bundles expresses as $$\Phi(z)=Res_{\ep}\Phi(z,\ep).$$
We call this by the same letter implying that the $z,w$
parameterize cotangent vectors while $\ep,\eta$ are respective for
the residue at schematic point. For this expression we obtain the
following Poisson brackets:
{\Lem \label{r-for-phiz}
\begin{equation}\label{PB7}
  \left\{\Phi(z)\otimes\Phi(w)\right\}=\left|[\Phi(z)\otimes
  1,R]\delta_w^z\right|_{-\{z,w\}}=\left|[\Phi(z)\otimes 1
  +1\otimes\Phi(w),R]\delta_w^z\right|_{w}.
\end{equation}
}
{\bf Proof~} The first equality follows from:
$$\{\Phi(z)\otimes\Phi(w)\}=Res_{\ep,\eta}\{\Phi(z,\ep)\otimes\Phi(w,\eta)\}=
\left|\left[Res_{\ep}\Phi(z,\ep)\otimes 1
,R\right]\delta_w^z\right|_{-\{z,w\}} $$ where we have used
(\ref{PB6}). The rest is proved as in lemma \ref{l1} $\square$

\subsection{Commutativity of Hitchin hamiltonians}

{\Def The quantities  $H_k(z)=Tr \left|\Phi^k(z)\right|_z$ are
called the Hamiltonians of our system.}
\\

The following theorem is the precursor of the integrability.
\begin{Th} The quantities $H_k(z)$ Poisson commute
$$\left\{H_k(z),H_m(w)\right\}=0.$$
\end{Th}
{\bf Proof~ }We use the linearity of Poisson brackets, the
$\left|...\right|_{w}$ operation and the Leibnitz rule: $$
\left\{\left|Tr\Phi^k(z)\right|_{z},\left|Tr\Phi^m(w)\right|_{w}\right\}=
\left|\left\{Tr\Phi^k(z),Tr\Phi^m(w)\right\}\right|_{w,z}=$$
\begin{equation} \label{com1}
Tr_{1,2}\left|\sum_{i=0,j=0}^{k-1,l-1}\Phi^i(z)\otimes\Phi^j(w)\left|[\Phi(z)\otimes
1+1\otimes
\Phi(w),\delta_{w}^{z}R]\right|_{w}\Phi^{k-1-i}(z)\otimes
\Phi^{l-1-j}(w)\right|_{w, z}
\end{equation}
We show now that interior $|...|_{w}$ is tautological. It is
always true for function $f(w), g(w)$ with only negative powers in
$w$ that $$\left|f(w)\left|g(w)\right|_{w}\right|_{w}=
\left|f(w)g(w)\right|_{w}.$$
 Applying this one obtains
$$(\ref{com1})=(kl)*Tr_{1,2}\left|\Phi^{i-1}(z)\otimes\Phi^{l-1}(w)
[\Phi(z)\otimes 1+1\otimes \Phi(w),\delta_{w}^{z}R]\right|_{w,z}$$
which is zero. $\blacksquare$

{\Cor
The Hitchin's Hamiltonians on the cotangent bundle to the moduli space
of vector bundles on curves Poisson commute for the case
of our singular curves also. And so they form the integrable system.}

This follows from the fact that hamiltonian reduction preserves the Poisson
bracket of invariant functions. Functions
$H_k(z)$ Poisson commute on the space of
pairs $\L(\ep),\Phi(\ep)$ as we just proved, they are invariant with
respect to the action of $GL(n).$ The cotangent bundle to the moduli space
of vector bundles on our singular curves can be obtained as hamiltonian
reduction from the space of pairs  $\L(\ep),\Phi(\ep)$ (see theorem
\ref{red-th}). And finitely Hitchin hamiltonians commute on the reduced space.
The calculation that the number of independent hamiltonians
is the half of the dimension of the phase space is
the same as in the case of nonsingular curves.

\section{Discussion and relations with other works}

A sort of similar constructions appears in the context of Beauville system
\cite{Bea,Mark} which is an integrable system on the space of rational matrices.
We have to mention that the specific choice of orbits in Beauville approach
corresponds to Hitchin system on singular curves (see \cite{T1}, examples
1,2). There will be interesting to generalize the way to choose orbits
for more complicated singularities considered in the present paper. Another
important analysis was effectuated in \cite{Chern} where it was considered
the limit procedure to obtain higher order poles in the Lax operator as a
fusion of simple poles and it was obtained an elliptic analog of the Lax
operator with double pole.

We should also remark that in \cite{Boalch}, \cite{BoalchBiq}
there were obtained some infinite-dimensional hamiltonian quotient
constructions do describe the finite-dimensional spaces of connections
with poles.
It would be interesting to work out analogous constructions for our
phase spaces this may give immediate proof of integrability
of some of our systems.

\end{document}